\documentclass[floatfix,groupedaddress,twocolumn,showpacs,amsmath,prx,aps,amssymb,longbibliography,nofootinbib]{revtex4-2}

\usepackage{graphicx}
\usepackage{xcolor}
\usepackage{lipsum}
\usepackage{overpic}
\usepackage[english]{babel}
\usepackage{amsmath,amsfonts,amssymb,float}
\usepackage[titletoc,title]{appendix} 
\usepackage{soul}
\usepackage{mathtools}
\usepackage{bbold}

\usepackage[unicode]{hyperref}
\hypersetup{
	unicode=true, 
	a4paper=true,
	plainpages=false,
	colorlinks=true,
	linkcolor=blue,
	citecolor=blue,
	filecolor=blue,
	urlcolor=blue
}
\newcommand{\pol}{\hat{\bf e}}
\newcommand{\rv}{{\bf r}}

\newcommand{\eo}{\epsilon_0}
\newcommand{\beq}{\begin{equation}}
\newcommand{\eeq}{\end{equation}}
\newcommand{\bea}{\begin{eqnarray}}
\newcommand{\eea}{\end{eqnarray}}
\newcommand{\BEQAL}{\begin{align}}
\newcommand{\EEQAL}{\end{align}}
\newcommand{\EQREF}[1]{Eq.~(\ref{#1})}

\newcommand{\comment}[1]{{}}

\newcommand{\<}{\langle}
\renewcommand{\>}{\rangle}

\renewcommand{\[}{\left[}
\renewcommand{\]}{\right]}

\newcommand{\commentout}[1]{{}}

\begin{document}

\title{Superradiant phase transition in a large interacting driven atomic\\ ensemble in free space}
\author{Janne Ruostekoski}
\affiliation{Department of Physics, Lancaster University, Lancaster, LA1 4YB, United Kingdom}
\date{\today}

\begin{abstract}
Atomic ensembles strongly interacting with light constitute rich quantum-optical many-body systems, with the potential for observing cooperative effects and dissipative nonequilibrium phase transitions.
We theoretically analyze the conditions under which a driven atomic ensemble in free space, characterized by strong dipole-dipole interactions and large spatial extent, can undergo a superradiant phase transition, also known as cooperative resonance fluorescence. In an atomic array, stationary states that conserve the collective pseudospin exhibit completely cooperative decay and undergo a second-order phase transition in the large atom number limit. In contrast, decay mechanisms on longer timescales that fail to conserve pseudospin can lead to discontinuous first-order phase transition at a critical finite atom number, disrupting cooperation despite sharing many similar observable characteristics. A hallmark of the superradiant phase transition is an abrupt shift from total light reflection off the atoms to rapidly increasing transmission, accompanied by significant quantum fluctuations, as a function of light intensity.
\end{abstract}

\maketitle

\section{Introduction}

Recent years have seen a dramatic resurgence in interest in collective optical interactions between atoms, stimulated by enhanced experimental imaging capabilities of cold atoms at high densities~\cite{Bender2010,BalikEtAl2013,Pellegrino2014a,Jenkins_thermshift,Jennewein_trans,Dalibard_slab,Saint-Jalm2018,Rui2020,Ferioli21,Srakaew22}. This has renewed focus on foundational models of cooperative phenomena, such as the Dicke model~\cite{Dicke54} for superradiance~\cite{Svidzinsky08,wilkowski,Araujo16,Roof16,Solano_super,Sutherland2017,Okaba2019,Robicheaux2021,Orioli2022,Masson2022,Asselie2022,Malz2022,Rubies-Bigorda2023,Liedl2024}. While the emission in Dicke superradiance, with the intensity proportional to the square of the atom number $N^2$, is essentially a classical effect of phased dipoles, its extension to a driven case---\emph{cooperative resonance fluorescence} (CRF)---reveals intricate quantum behaviors and phase transition phenomena~\cite{Senitzky1972,Agarwal1977,Drummond1978,Walls1978,Narducci1978,Puri1979,Carmichael1980,Drummond1980b}. In CRF, an idealized system of noninteracting two-level atoms are confined to volume smaller than a cubed resonant wavelength, such that their identical response to light can be represented by a collective pseudospin operator $\hat S_\pm=\sum_j \hat\sigma^\pm_j$ and $\hat S_z=\sum_j \hat\sigma^z_j$, with $\hat S_\pm=\hat S_x\pm i \hat S_y$, where $\hat\sigma^+_j=|e\>\<g|$ is the raising spin operator for atom $j$.  This configuration conserves the total pseudospin $\<\hat S^2\>$, or the radius of the Bloch sphere, and decays through a completely cooperative mechanism, which in less idealized scenarios is limited by short coherence times. The CRF supports two nonequilibrium phases: At weak driving, fluorescence is classical from an induced collective dipole that attenuates the incident field. At high intensities, the system undergoes a second-order phase transition where, in the limit $N\rightarrow\infty$, the excited level sharply saturates, the mean dipole moment vanishes, and quantum fluctuations increase significantly.

Inspired by a recent experimental report on the observation of the superradiant phase transition~\cite{Ferioli2023}, traditionally referred to as CRF, in a dilute atom cloud in free space, we develop a simple and tractable theoretical model to explore this phase transition within a spatially extended atomic ensemble experiencing strong dipole-dipole (DD) interactions in free space. We discover that, in a planar array of closely spaced atoms in a superradiant mode with a broad resonance linewidth, the system can conserve a macroscopic pseudospin on short timescales, leading to complete cooperative decay and a second-order phase transition. A hallmark of this phase transition is an abrupt shift in the array's properties from a perfect mirror to rapidly increasing light transmission, accompanied by enhanced quantum fluctuations.
Both the transition point and the scattered light intensity depend on the self-interaction resonance linewidth leading to different scaling with the atom number compared with the idealized noninteracting system confined to volume with
dimensions much smaller than a wavelength.
The counterintuitive result that strong DD interactions enable CRF can be attributed to delocalized excitations that, paradoxically, enhance the indistinguishability of the atoms.

When the pseudospin is not conserved, the behavior of single-atom observables at large atom numbers can still, surprisingly, approximate the sharp functional dependence typical of CRF. At high densities, the decay mechanism that fails to conserve the pseudospin may instead undergo a first-order phase transition at a critical atom number, indicative of optical bistability~\cite{bonifacio1976,Bonifacio1978,Carmichael1977,Agrawal79,Drummond1980,Drummond1981, Parmee2021}. We demonstrate that sufficiently weak position fluctuations do not prevent bistability, and how even stronger fluctuations can still preserve other distinct characteristics of single-atom observables. Independently to our work, the CRF experiment~\cite{Ferioli2023} is theoretically analyzed in Refs.~\cite{agarwal2024,goncalves2024}.  

\section{Atom-light interactions}

We study $N$ two-level atoms in a 2D square array with lattice constant $a$ and unit filling. Light mediates strong DD interactions through multiple scattering. The quantum-optical properties of such planar arrays have recently been reviewed in Ref.~\cite{Ruostekoski2023}. We first analyze the atomic coherences ${\rho}^{(j)}_{ge}=\<\hat\sigma^-_j\>$ and the excited level populations $ {\rho}^{(j)}_{ee}$ of atom $j$ at $\rv_j$ employing the semiclassical approximation where quantum fluctuations between different atoms~\cite{Bettles2020} are ignored (the ground state populations are derived from $\rho_{gg}^{(j)}= 1- \rho_{ee}^{(j)}$). 
The dynamics in the rotating-wave approximation are then given by 
\begin{subequations}\label{Eq:SpinEquations2LS}
\begin{align}
	\dot{\rho}^{(j)}_{ge}=&\left(i\Delta-\gamma\right) {\rho}^{(j)}_{ge}-i(2{\rho}^{(j)}_{ee}-1)\mathcal{R}_{\rm eff}^{(j)},\label{Eq:Coherence}\\
	\dot{\rho}^{(j)}_{ee}=&-2\gamma {\rho}^{(j)}_{ee}+2\text{Im}[(\mathcal{R}_{\rm eff}^{(j)})^*{\rho}^{(j)}_{ge}],
	\label{Eq:Excitations}
\end{align}
\end{subequations}
where $\gamma= {\cal D}^2 k^3/(6\pi\hbar\eo)$ is the single-atom resonance linewidth, ${\cal D}$ the dipole matrix element, with the atomic dipole ${\bf d}= {\cal D} \hat{\textbf{e}}_d $, and $k=2\pi/\lambda$ the resonance wavenumber.  Effective Rabi frequencies~\cite{Parmee2020}, 
\beq
\label{Eq:EffectiveRabi}
	\mathcal{R}_{\rm eff}^{(j)} = \mathcal{R}^{(j)} +  \xi \sum_{\ell\neq j} \mathsf{G}^{(j\ell)}_{dd}\rho_{ge}^{(\ell)}, \quad \xi=\frac{{\cal D}^2}{\hbar\eo},
\eeq
drive atom $j$
and include the contributions from the incident field Rabi frequency $\mathcal{R}^{(j)}={\cal D}\hat{\textbf{e}}_d\cdot\boldsymbol{\mathcal{E}}{}^+(\textbf{r}_j)/\hbar $ and from the scattered fields from all the other atoms at $\rv_\ell$ through the dipole radiation kernel $\mathsf{G}(\textbf{r})$~\cite{Jackson} 
\beq
{\sf G}_{\nu\mu}({\bf r}) =
\left[ {\partial\over\partial r_\nu}{\partial\over\partial r_\mu} -
\delta_{\nu\mu} \boldsymbol{\nabla}^2\right] {e^{ikr}\over4\pi r}
-\delta_{\nu\mu}\delta({\bf r})
\label{eq:GDF}
\eeq
in $\mathsf{G}^{(j\ell)}_{dd}=\hat{\textbf{e}}_d^*\cdot \mathsf{G}(\textbf{r}_j-\textbf{r}_\ell)\hat{\textbf{e}}_d$. The positive frequency component of the light amplitude $\boldsymbol{\mathcal{E}}{}^+(\textbf{r}_j)$ and ${\rho}^{(j)}_{ge}$ are expressed as slowly varying amplitudes where the rapid oscillations at the laser frequency have been factored out.
The second term in Eq.~\eqref{Eq:EffectiveRabi}  can induce recurrent scattering and strong DD interactions. Omitting this term results in the standard optical Bloch equations.
The semiclassical theory of \EQREF{Eq:SpinEquations2LS} for planar arrays was found in Ref.~\cite{Bettles2020} to agree well with the quantum solution, particularly when exciting collective modes with broad superradiant resonances, as we consider here. However, quantum effects are crucial in the dynamics of phase transitions.

We now develop a simple model to describe the steady states of Eqs.~\eqref{Eq:SpinEquations2LS} and~\eqref{Eq:EffectiveRabi}. Assuming phase-uniform incoming light that is normally incident to the array, we treat each atom as equally driven, with $\mathcal{R}^{(j)}=\mathcal{R}$. We disregard edge effects, an approximation that becomes accurate for large arrays.  We concentrate on stable steady states with a uniform phase profile that directly couple to the incident light. Consequently, we set $ {\rho}^{(j)}_{ee}= {\rho}_{ee}$ and $ {\rho}^{(j)}_{ge}= {\rho}_{ge}$, resulting in uniform $\mathcal{R}_{\rm eff} = \mathcal{R}+ {\rho}_{ge} \xi \sum_{\ell\neq j}\mathsf{G}^{(j\ell)}_{dd} = \mathcal{R}+ 
(\tilde\Omega+i\tilde\gamma) {\rho}_{ge}$, where  $\tilde\Omega$ and $\tilde\gamma$ represent the real and imaginary components, respectively, of the radiation kernel  $\tilde\Omega+i\tilde\gamma =  \xi \sum_{\ell\neq j}\mathsf{G}^{(j\ell)}_{dd}$~\cite{Parmee2020}. In the absence of edge effects, the lattice is considered translationally invariant, rendering the position of site $j$ arbitrary. The steady-state solutions of \EQREF{Eq:SpinEquations2LS} now read
\begin{subequations}\label{Eq:EOMSolns}
	\begin{align}
	\rho_{ge} &={\cal R}_{\rm eff}\frac{-\Delta+i\gamma}{\Delta^2+\gamma^2+2|{\cal R}_{\rm eff}|^2},\label{Eq:Coherencess}\\
	\rho_{ee} &=\frac{|{\cal R}_{\rm eff}|^2}{\Delta^2+\gamma^2+2|{\cal R}_{\rm eff}|^2} \label{Eq:Excitationsss}.
	\end{align}
\end{subequations}
Analogous uniform excitations were also observed experimentally in the transmission of light through an atomic planar array, conducted in the low light intensity (LLI) limit in Ref.~\cite{Rui2020}. In practise, possessing a large array size is not essential for the effectiveness of the approximation 
\EQREF{Eq:EOMSolns}; it suffices that the phase profiles are uniform. In finite arrays, the coupling to phase-uniform modes can be significantly enhanced by adjusting the width of a Gaussian laser beam to match the lattice size~\cite{Facchinetti16}. A detailed analysis of mode matching errors was carried out in Ref.~\cite{Manzoni18}.

Using Eq.~\eqref{Eq:Coherencess} and the expression for ${\cal R}_{\rm eff}$, we obtain the relationship between $\mathcal{R}$ and ${\cal R}_{\rm eff}$ depending on the cooperativity parameter $C = (\tilde{\Omega}+i\tilde{\gamma})/[2(\Delta+i\gamma)]$~\cite{Parmee2020},
\begin{equation}\label{Eq:yequation}
\mathcal{R}={\cal R}_{\rm eff}+{\cal R}_{\rm eff}\frac{2C(\Delta^2+\gamma^2)}{\Delta^2+\gamma^2+2|\mathcal{R}_{\rm eff}|^2}.
\end{equation}
Taking  the absolute value of both sides of Eq.~\eqref{Eq:yequation} yields a cubic equation in $|\mathcal{R}_{\rm eff}|^2$, which may have one or two dynamically stable solutions---indicating optical bistability. 

\subsection{Low light intensity excitation eigenmodes}

Even beyond the LLI regime, the collective atom behavior can be described by the properties of underlying LLI collective radiative excitation eigenmodes.
In the LLI limit~\cite{Ruostekoski1997a}, \EQREF{Eq:SpinEquations2LS} simplifies to \EQREF{Eq:Coherence} with $\rho_{ee}=0$. The resulting dynamics of coupled linear dipole oscillators are expressed as: 
\beq
\dot{\rho}^{(j)}_{ge}=i\sum_\ell ({\cal H}_{j\ell}+\delta{\cal H}_{j\ell}) {\rho}^{(\ell)}_{ge}+ i \mathcal{R}^{(j)},
\eeq
where the matrix $\delta\mathcal{H}_{j\ell}$ includes $\Delta$ on the diagonal and 
\beq
\mathcal{H}_{j\ell} =i\gamma\delta_{j\ell}+ \xi \mathsf{G}^{(j\ell)}_{dd} (1-\delta_{j\ell})
\eeq
provides the collective excitation eigenmodes~\cite{Rusek96, Jenkins_long16}. The eigenvalues $\delta_j+i\upsilon_j$ yield the LLI collective line shift $\delta_j$ and collective linewidth $\upsilon_j$. For the uniform LLI eigenmode, the eigenvalue $\eta$ satisfies~\cite{Ruostekoski2023} $\mathcal{H}\rho_{ge}=\eta\rho_{ge}$, or 
\beq
\mathcal{H}\rho_{ge}=(i\gamma+ \xi \sum_{\ell\neq j} \mathsf{G}^{(j\ell)}_{dd} ) \rho_{ge}=[\tilde\Omega+i(\gamma+\tilde\gamma)] \rho_{ge}, 
\eeq
with the line shift $\tilde\Omega$ and linewidth $\upsilon=\gamma +\tilde\gamma$. Both $\tilde\Omega$ and $\tilde\gamma$ can be numerically calculated for finite arrays. These parameters critically influence the behavior of the effective field ${\cal R}_{\rm eff}$ in \EQREF{Eq:yequation} through the cooperativity parameter  $C$.

\subsection{Collective $\<\hat S^2\>$ breaking interactions}

\begin{figure}
	\hspace*{0cm}
	\includegraphics[width=0.98\columnwidth]{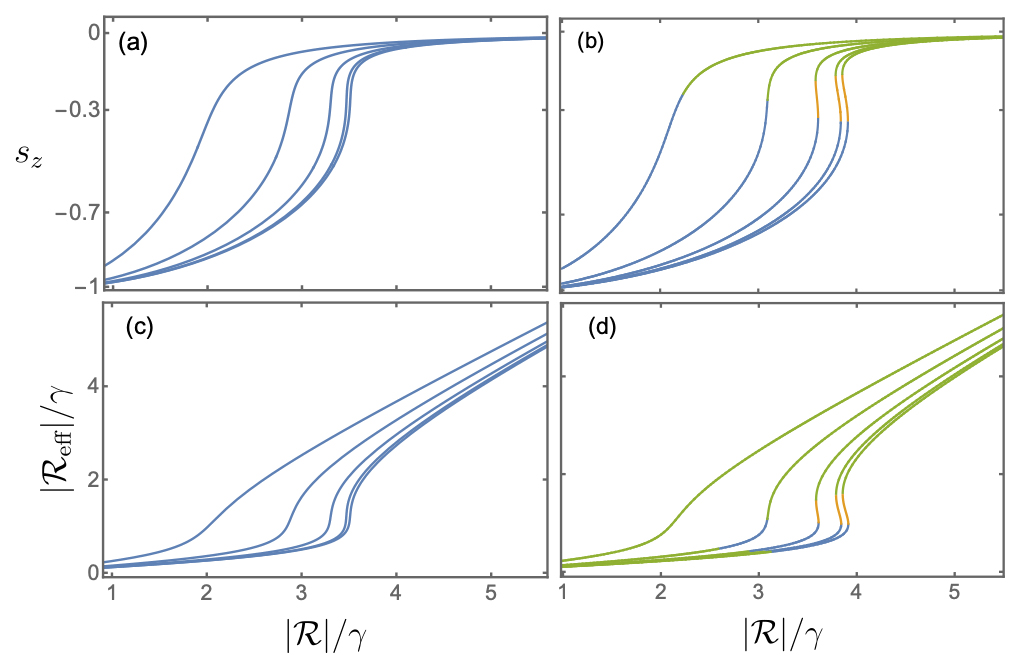}
	\vspace{-0.7cm}
	\caption{The atom number $N$ dependence of (a,b) the collective atomic pseudospin component $s_z$ and (c,d) the effective Rabi frequency $|{\cal R}_{\rm eff}|$ driving each atom in the array as a function of the Rabi frequency of the incident driving $|{\cal R}|$ for lattice constants (a,c) $a/\lambda=0.17$ and (b,d) $0.16$. The curves from top to bottom $N=4,9,16,25,36$ reaching the asymptotic large $N$ value. (b,d) First-order phase transition and optical bistability above a critical $N$ are observable in the last three curves; (a,c) no phase transition.}
	\label{Fig:bista}
\end{figure}

\begin{figure}
	\hspace*{0cm}
        \includegraphics[width=0.48\columnwidth]{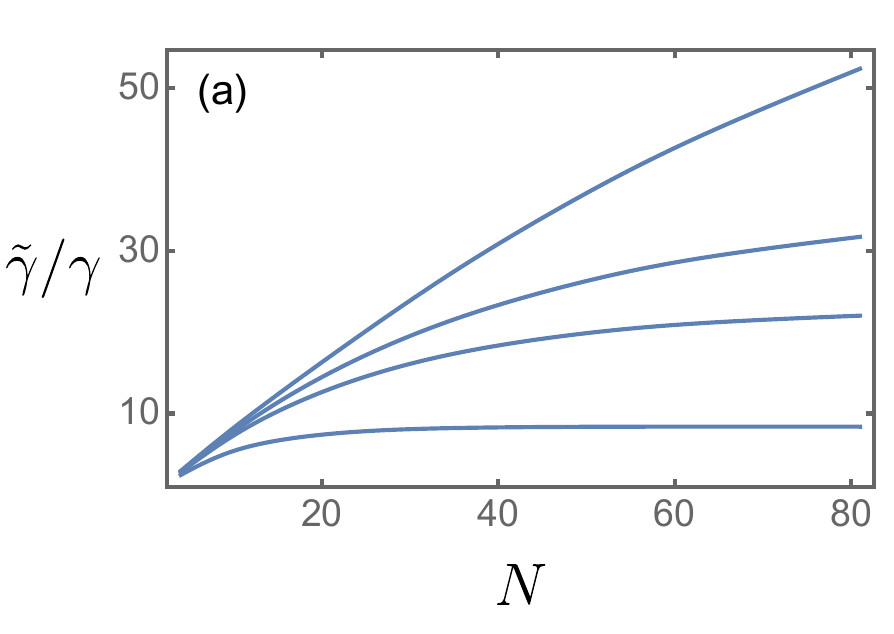}
        \includegraphics[width=0.48\columnwidth]{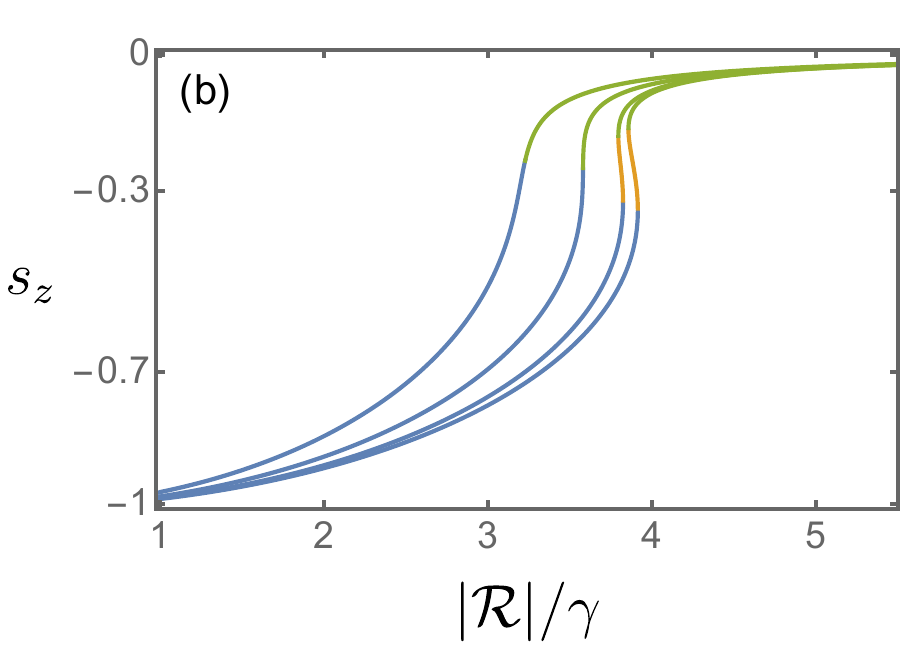}
	\vspace{-0.4cm}
	\caption{(a) The atom number $N$ dependence of the collective self-interaction linewidth $\tilde\gamma$. The curves from top to bottom for the lattice constants: $a/\lambda=0.05$, 0.08, 0.10, and 0.16. (b) The effect of position fluctuations of atoms on the bistability phase transition in the large $N$ limit for $a=0.16 \lambda$ shown in Fig.~\ref{Fig:bista}(b). The curves from top to bottom for the atom density root-mean-square widths $\eta/a$ at each site: 0.3, 0.2, 0.1, and 0. The phase transition is observable at $\eta/a=0$ and 0.1.}
	\label{Fig:fluc}
\end{figure}

Next we solve \EQREF{Eq:SpinEquations2LS} for varying atom numbers $N$ by solving the cubic equation for $|\mathcal{R}_{\rm eff}|^2$ derived from \EQREF{Eq:yequation}. The dependence on $N$ arises through the cooperativity parameter $C$, obtained by numerically calculating $\tilde\Omega(N)$ and $\tilde\gamma(N)$ for different $N$. In Fig.~\ref{Fig:bista}, we show $s_z=2\rho_{ee}-1$ and $|{\cal R}_{\rm eff}|$ for lattice spacings $a/\lambda=0.17$ and $0.16$, and examples of $\tilde\gamma$ as a function of $N$ are displayed in Fig.~\ref{Fig:fluc}(a).
As $N$ increases, $s_z$ develops a more pronounced concave curve, reaching a sharp inflection point near saturation---traits also characteristic of CRF. Initially, for increasing intensities, the incident light is dampened by the induced atomic polarization, resulting in weak $|{\cal R}_{\rm eff}|$,  until reaching a threshold intensity where $|{\cal R}_{\rm eff}|$ begins to increase rapidly. The sharpening of these curve profiles results from the broadening linewidth $\gamma+\tilde\gamma$. However, no phase transition occurs at $a=0.17\lambda$. The scenario changes at $a=0.16\lambda$ where, beyond a critical atom number, $|{\cal R}_{\rm eff}|$ exhibits three solutions, indicating optical bistability and representing a discontinuous first-order phase transition. 
Two of the solutions are stable and the one with a negative gradient unstable. Analogous to optical cavity bistability~\cite{Bonifacio1978}, the two stable branches in Fig.~\ref{Fig:bista} are referred to as cooperative (low $s_z$) and single-atom (high $s_z$) solutions~\cite{Parmee2021}.
The bistability is intrinsic, generated by DD interactions in free space in the absence of a cavity. 
Theoretical studies suggest that such interaction-mediated bistable behavior can be generic in cold-atom systems with a variety of short- and long-range interactions~\cite{Lee2011,Sibalic2016,Parmee2018}, and it has been experimentally observed in Rydberg atoms in the microwave regime~\cite{Carr2013}. Small lattice constants $a/\lambda\simeq0.08$ and 0.17 at magic wavelengths that generate strong DD interactions are achievable with Sr~\cite{Olmos13,Ballantine22str} and Yb~\cite{Beloy12} atoms, respectively.

As the atom positions are encoded in the collective linewidths $\gamma+\tilde\gamma$ and line shifts $\tilde\Omega$, we examine the impact of weak position fluctuations on the phase transition by calculating the changes in $\gamma+\tilde\gamma$ and $\tilde\Omega$ due to disorder. We solve the coupled LLI equations for the eigensystem ${\cal H}$, randomly sampling atom positions around each lattice site from Gaussian density distributions with a root-mean-square width $\eta$~\cite{Jenkins2012a}. By ensemble-averaging over many stochastic realizations, we obtain the expectation values of $\<\tilde\gamma\>$ and $\<\tilde\Omega\>$. With sufficiently weak fluctuations, the relevant mode largely retains its phase uniformity. In deep optical lattice potentials, $\eta$ can be related to the lattice height in terms of lattice photon recoil energies $s$, with $\eta/a=s^{-1/4}/\pi$~ \cite{Morsch06}. Intriguingly, as demonstrated in Fig.~\ref{Fig:fluc}, while bistability persists under weak fluctuations, stronger fluctuations may eliminate bistability yet preserve the sharp features of the $s_z$ curve, reminiscent of CRF.

\subsection{Collective $\<\hat S^2\>$ conserving interactions}

As we analyze \EQREF{Eq:SpinEquations2LS} for broad superradiant resonances, where the linewidths satisfy $\upsilon \gg\gamma$,  we can neglect the terms proportional to $\gamma$ for observation times shorter than $1/\gamma$. Consequently, we reformulate  \EQREF{Eq:SpinEquations2LS} for uniform excitations using scaled collective pseudospin expectation values $s_z=2\<\hat S_z\>/N=2\rho_{ee}-1$ and $s_-=2\<\hat S_-\>/N=2\rho_{ge}$: 
\begin{subequations}\label{Eq:SpinEquations2LS_B}
\begin{align}
	\dot{s}_{-} & = i\Delta {s}_{-}-i  s_z[2\mathcal{R}+(\tilde\Omega+i\tilde\gamma) {s}_{-}],\label{Eq:CoherenceB}\\
	\dot{s}_z & = i \mathcal{R}(s_+ - s_-)-\tilde\gamma |s_-|^2,
	\label{Eq:ExcitationsB}
\end{align}
\end{subequations}
where, without the loss generality, we have taken $\mathcal{R}$ to be real.
We discover that $\frac{d}{dt} (s_z^2+  |s_-|^2) =0$, indicating that  the collective pseudospin $s^2=s_z^2+ s_x^2+s_y^2\leq1$ is conserved for the macroscopic spin  $S=N/2$. Remarkably, this dynamic closely mirrors the idealized CRF observed in entirely noninteracting atoms confined to a volume smaller than a cubed resonant wavelength. Analyzing the steady-state solutions where $\dot s_-=\dot s_z=0$ at the collective resonance $\Delta=s_z\tilde\Omega$, we derive from \EQREF{Eq:CoherenceB} that either $s_z=0$ or  $s_-=i\beta$, where 
$\beta= 2\mathcal{R}/\tilde\gamma$.

{\it The case $s_-=i\beta$:} From \EQREF{Eq:ExcitationsB} and the conservation equation for $s^2$, we derive $s_z=-\sqrt{s^2-\beta^2}$, where the positive sign before the square root does not yield a stable solution. This relationship holds for $\beta^2 \leq s^2\leq1$, encompassing the weak field limit where the excited level population vanishes ($s_z=-1$). 
As the incident light amplitude increases, the coherence $\rho_{ge}$ rises linearly until it reaches its maximum at $|\rho_{ge}|=|s|/2$, corresponding to the saturation point of the excited level at $\rho_{ee}=1/2$ ($s_z=0$). 
In this scenario, the increasing induced dipole $\rho_{ge}$ attenuates incident driving, resulting in a weak effective Rabi frequency driving each atom $\mathcal{R}_{\rm{eff}}= i \mathcal{R}\tilde\Omega/\tilde\gamma$. Notably, at the single-atom resonance ($\Delta=0$) $\mathcal{R}_{\rm{eff}}=0$ exactly cancels out. 
For atomic arrays, the collective behavior can be observed in light transmission experiments, analogous to Refs.~\cite{Rui2020,Srakaew22} (see Appendix~\ref{sec:app}).
At the collective resonance the array  acts as a perfect mirror $r\simeq-1$, for $\tilde\gamma\gg\gamma$ [\EQREF{Eq:Transmission2}]. Intriguingly, this regime of perfect reflection spans the entire range of $\rho_{ee}$ from zero to saturation, significantly extending the LLI observations of an atomic mirror~\cite{Rui2020}.

{\it The case $s_z=0$:} From the conservation of $s^2$ and \EQREF{Eq:ExcitationsB}, we derive $s_-= i s^2/\beta \pm\sqrt{s^2 - s^4/\beta^{2}}$ for $\beta^2\geq s^2$. As $\beta$ increases, while $s_z=0$ remains saturated, the magnitude of $|\rho_{ge}|$ no longer escalates or offsets the increasing incident light, resulting in rapid growth in $\mathcal{R}_{\rm{eff}}$. The power reflection $|r|^2 \simeq\beta^{-2}$ begins to decrease as $ \mathcal{R}$ rises. Correspondingly, the transmission $|t|^2\simeq 1- \beta^{-2}$ increases.
Different factorizations can lead to different mean-field solutions. 
Analogous to CRF, these solutions are not stable~\cite{Drummond1978}; rather, small perturbations result in periodic oscillations around the stationary state.

\begin{figure}
	\hspace*{0cm}
	\includegraphics[width=0.48\columnwidth]{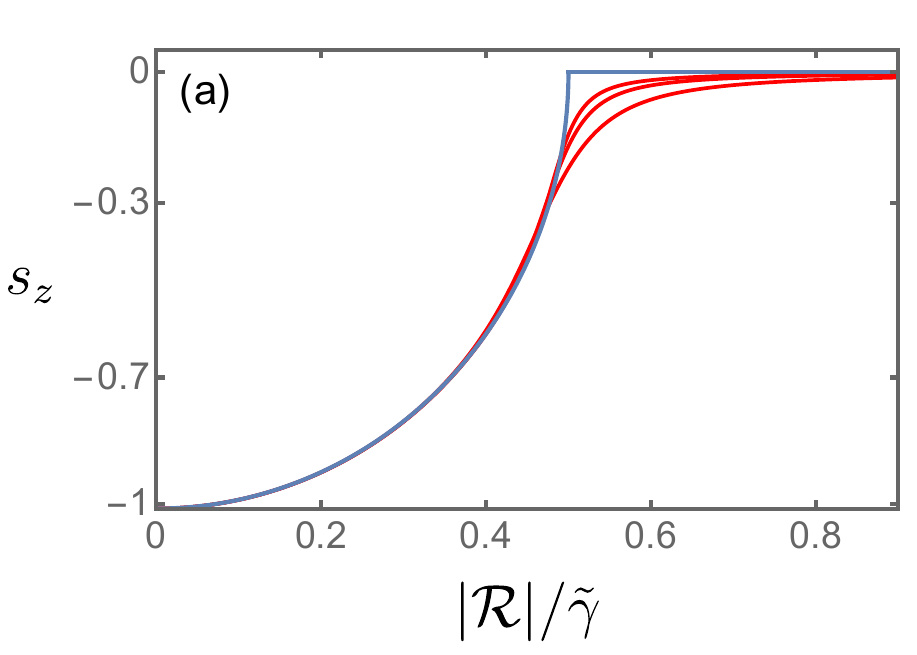}
	\includegraphics[width=0.48\columnwidth]{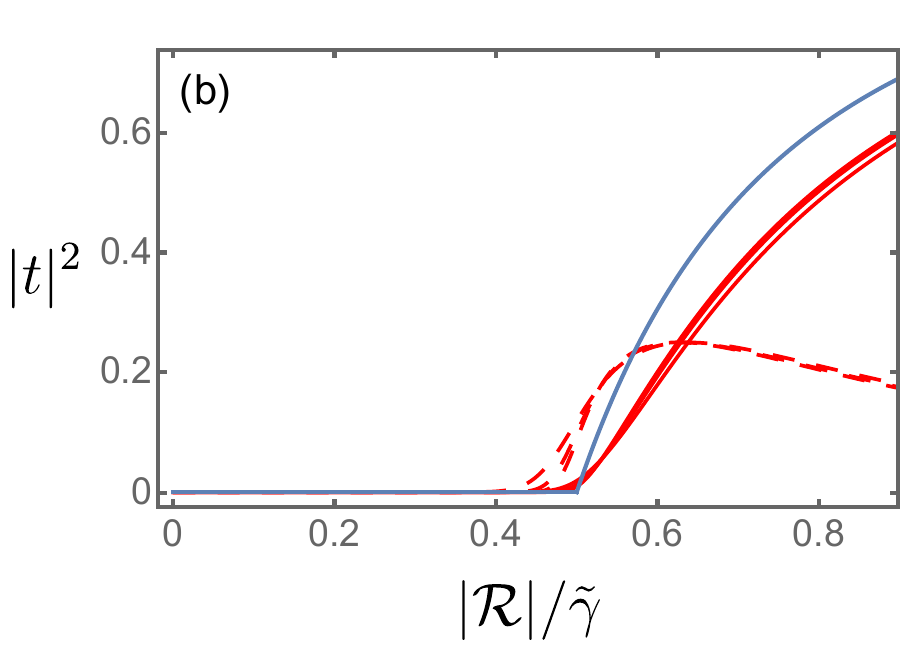}
	\vspace{-0.2cm}
	\caption{Cooperative resonance fluorescence. (a) The collective atomic spin $s_z$ and (b) power transmission coefficient $|t|^2$ for coherently scattered light through the atom array display a second-order phase transition as a function of ${\cal R}$ scaled by the collective self-interaction linewidth $\tilde\gamma$.  In (a), the curves from top to bottom:  the semiclassical solution, quantum solutions for $N=80$, 50, and 25. The quantum solutions asymptotically approach the semiclassical curve with discontinuous derivative. In (b), the perfect reflection is maintained at intensities all the way to saturation point, after which the perfect array reflection sharply transitions to rising transmission.
 The quantum solutions (the lower solid curves) asymptotically approach a curve distinct from the semiclassical result (top curve). The quantum fluctuations derived from $\<s_+ s_-\>-|\<s_+\>|^2$ are shown in the correspnding incoherent transmission contribution (dashed line).
}
	\label{Fig:quantum}
\end{figure}

We can formulate an approximate quantum master equation for the density matrix $\hat \rho$ by focusing only on the coupling to the phase-uniform collective mode, analogously to the approach used in  \EQREF{Eq:SpinEquations2LS_B}. Assuming the collective resonance, the quantum analog of \EQREF{Eq:SpinEquations2LS_B} reads
\beq
\label{eq:rhoeom}
\dot{\hat \rho} = i \big[2{\cal R} \hat S_x,\hat \rho\big] +\frac{2\tilde\gamma}{N} \left(
2\hat S_-\hat \rho\hat S_+ -\hat S_+ \hat S_-\hat \rho -\hat\rho\hat S_+ \hat S_-\right),
\eeq
where the bracket denotes a commutator. The relationship between Eqs.~\eqref{eq:rhoeom} and~\eqref{Eq:SpinEquations2LS_B} is then the same as the one between the quantum and semiclassical theories of CRF of noninteracting atoms confined to volume with dimensions much smaller than a wavelength~\cite{Agarwal1977,Drummond1978,Walls1978,Narducci1978,Puri1979,Carmichael1980,Drummond1980b}. 
The single-atom linewidth $\gamma$ of the CRF master equation is replaced in Eq.~\eqref{eq:rhoeom} by the scaled collective self-interaction linewidth of the atom array $2\tilde\gamma/N$. This modifies the dependence of the transition point on the atom number from $N/2$ to a slower-than-linear scaling with $\tilde\gamma/\gamma$ [see Fig.~\ref{Fig:fluc}(a)].
Although the solutions to Eq.~\eqref{eq:rhoeom}  have been thoroughly explored (also for more recent results see, e.g, Ref.~\cite{Hannukainen2018}) we can directly relate them to the power transmission coefficients $|t|^2$ of coherently scattered light through the atomic array (Appendix~\ref{sec:app}). In Fig.~\ref{Fig:quantum}, the rescaled steady-state solutions of \EQREF{eq:rhoeom} for the spin expectation values and transmission are shown for varying $N$ alongside the mean-field solutions of \EQREF{Eq:SpinEquations2LS_B}. 
Both cases converge at a critical transition point where the excited state saturates, and perfect array reflection sharply transitions to increased transmission.

For $\<s_z\>$ in Fig.~\ref{Fig:quantum}(a), the quantum solution smoothly approaches the mean-field solution as $N\rightarrow\infty$, marking a clear second-order phase transition. However, the quantum solution $\<s_x\>=0$ significantly differs from the mean-field theory~\cite{Carmichael1980}, with discrepancies also noted for $\<s_y\>$. These differences impact $|t|^2$, shown in Fig.~\ref{Fig:quantum}(b), where the quantum solutions for increasing $N$ asymptotically approach a curve distinct from the mean-field predictions, indicating strongly enhanced quantum fluctuations. The expression $|\<s_+\>|^2+\<s_z\>^2$ is no longer conserved, unlike $\<s_+ s_-\>+\<s_z^2\>$~\cite{Carmichael1980}, as demonstrated in Fig.~\ref{Fig:quantum}(b) by the component of incoherently transmitted light originating from $\<s_+ s_-\>-|\<s_+\>|^2$ (Appendix~\ref{sec:app}). Fluctuations in light transmission are a strong signature of significant quantum effects beyond the mean-field theory.

\subsection{Scattered intensity}

We now examine how the scattered light intensity depends on the atom number  $N$. From the single-atom contributions in Eqs.~\eqref{Eq:EOMSolns} we find that the coherent scattering due to $\rho_{ge}$ is dominant at low intensities, whereas at high intensities, incoherent scattering from $\rho_{ee}$ becomes prevalent. The general expression for coherent scattering rate of photons, integrated over a closed surface enclosing the source, is given by (see Appendix~\ref{sec:app})
\beq\label{eq:cohscatrateformula} 
n_c =  2\gamma  \sum_{j} \langle \hat{\sigma}_{j}^+\>\< \hat{\sigma}_{j}^-\rangle  + 2 \xi \sum_{j\ell (j\neq\ell)} {\rm Im}[\mathsf{G}^{(j\ell)}_{dd}] \langle \hat{\sigma}_{j}^+\>\< \hat{\sigma}_{\ell}^-\rangle.
\eeq
For the phase-uniform mode we directly obtain 
\beq \label{eq:nratecoh}
n_c=2N(\gamma+\tilde\gamma)|\rho_{ge}|^2, 
\eeq
since $\langle \hat{\sigma}_{j}^\pm\>$ can be taken outside the sum. More generally for the array, the correlations due to DD interactions alter the typical scaling from $N^2$ to $N\tilde\gamma/\gamma$ where $\tilde\gamma$ initially increases with small $N$ but quickly begins to saturate. For instance, for the lower (cooperative) branch of bistable solutions at weak incident driving ${\cal R}_{\rm eff}\simeq{\cal R}/(2C+1)$~\cite{Parmee2021}, and assuming $\tilde\gamma\gg {\cal R},\tilde\Omega$, we derive $n_c\simeq 2 N {\cal R}^2/\tilde\gamma$. Numerically, for different incident fields and lattice constants, we typically find an intensity scaling of $\propto N^{1+\alpha}$ where $\alpha$ is a small positive or negative number $0.1\alt |\alpha|\alt 0.4$. 
An analogous formula applies for incoherent scattering, with the necessary adjustments in the correlation functions (Appendix~\ref{sec:app}). 
For the upper (single-atom) branch of bistable solutions at high intensities, the single-atom incoherent scattering dominates as ${\cal R}_{\rm eff}\simeq{\cal R}$~\cite{Parmee2021}, resulting in the intensity being proportional to $N$. In the case of CRF from \EQREF{eq:rhoeom}, the two-atom quantum correlations become significant, with the asymptotic scaling of the scattered intensity again $\propto N \tilde\gamma/\gamma$, diverging from the noninteracting result $\propto N^2$~\cite{Carmichael1980,Drummond1980b}.  
\section{Concluding remarks}

We have demonstrated how DD interactions can lead to pronounced behaviors of saturation and effective driving field curves, phase transitions, and significant quantum fluctuations beyond the mean-field theory. 
Contrary to the gradual change observed in the second-order phase transition, where the sharp transition point is reached as $N\rightarrow \infty$, the discontinuous first-order transition occurs at a finite critical $N$. A full quantum treatment would also affect the first-order phase transition. It is generally well understood that bistable behavior coincides with enhanced quantum fluctuations~\cite{Carmichael1986a,Olmos2014,Parmee2018,Rempe1991,Drummond1980,Drummond1981, Parmee2020,Alaeian2022,Leppenen2024}, resulting in bimodal quantum distributions 
that represent tunnelling between the two mean-field solutions.

Historically, the CRF phase transition has been regarded as a highly idealized theoretical model, posing significant experimental challenges.  Ideally, every atom should experience the same field exposure and no interactions, any deviation from which could severely limit the coherence time of the decay. These conclusions also align with the recent theoretical studies parallel to ours~\cite{agarwal2024,goncalves2024}. 
Our interacting lattice scheme faces similar challenges, although we discovered that strong DD interactions in regular arrays can enhance the necessary indistinguishability of the atoms through delocalization of excitations. In practical experiments, the atoms in the array should be prepared close to the steady states of the CRF model, achievable by modulating field, atom, or lattice parameters. For instance, the state $s_-=i\beta$, $s_z=-\sqrt{s^2-\beta^2}$ could be prepared from another driven pure steady state by adjusting the driving amplitude.

\acknowledgments{We acknowledge financial support from the UK EPSRC Grant No.\ EP/S002952/1.}

\appendix

\section{Scattered intensity and transmission}
\label{sec:app}

In atomic arrays, collective behavior can be observed in light transmission experiments~\cite{Rui2020,Srakaew22}.
For large atomic layers, coherent light transmission and reflection of a plane wave demonstrate behavior similar to that in 1D electrodynamics for light~\cite{Javanainen19}. The amplitude of coherently scattered light amplitude from a square array of uniformly excited atoms propagates as a collimated wave. In the limit of an infinite array, the scattered field, with the density of dipoles ${\cal D}/a^2$ in a phase-uniform excitation, almost perfectly emulates a plane wave~\cite{Ruostekoski2023} 
\beq
\eo \<\hat{\textbf{E}}{}^+_s(x)\> = \frac{ik\pol_d {\cal D}}{2a^2}\rho_{ge} e^{ik|x|}, 
\eeq
assuming the normal to the lattice aligns with the $x$-axis.
The resonance linewidth of the corresponding uniform collective LLI eigenmode derives from the imaginary part of the 1D dipole radiation kernel~\cite{Ruostekoski2023}
\begin{align}
\label{eq:planarwidth1}
\upsilon &=\gamma+\tilde\gamma =\gamma  +   \xi \sum_{\ell\neq j} {\rm Im}  \[ \mathsf{G}^{(j\ell)}_{dd} \] \nonumber\\
& =\frac{\xi}{2a^2}\lim_{x\rightarrow 0} {\rm Im}[ik e^{ik|x|}] = \frac{3\pi\gamma}{k^2 a^2}. 
\end{align}
Combining these two expressions yields the coherent transmission and reflection amplitude coefficients:
\begin{equation}\label{Eq:TransmissionAmplitude}
t=\frac{\int_{x>0}\hat{\textbf{e}}_d\cdot\<\hat{\textbf{E}}{}^+(x)\>d\Omega}{\int_{x>0}\hat{\textbf{e}}_d\cdot\boldsymbol{\mathcal{E}}{}^+(\textbf{r})d\Omega},\quad r=\frac{\int_{x<0}\hat{\textbf{e}}_d\cdot\<\hat{\textbf{E}}{}^+_s(x)\>d\Omega}{\int_{x>0}\hat{\textbf{e}}_d\cdot\boldsymbol{\mathcal{E}}{}^+(\textbf{r})d\Omega},
\end{equation}
where $t=1+r$ due to the symmetry, and we obtain
\begin{equation}\label{Eq:Transmission2}
r = i (\gamma+\tilde{\gamma})\frac{\rho_{ge}}{\mathcal{R}}.
\end{equation}
Here $\rho_{ge}$ can be evaluated from \EQREF{Eq:Coherencess}.

We express the total scattered intensity as a photon scattering rate over a surface $S$ that completely encloses the atoms.  
The rate is given by the integral of the scattered intensity per photon energy
\beq \label{eq:emissionrateapp}
n_s = \frac{1}{\hbar\omega} \int_{S} dS I_s = \frac{ 2\eo c}{\hbar\omega} \int_{S} dS \<  \hat{\textbf{E}}{}^-_s(\textbf{r}) \cdot\hat{\textbf{E}}{}^+_s(\textbf{r}) \>.
\eeq
The general expression can be evaluated~\cite{carmichael2000,Ruostekoski2023}:
\beq\label{eq:scatrateformula} 
n_s =  2\gamma  \sum_{j} \langle \hat{\sigma}_{j}^+ \hat{\sigma}_{j}^-\rangle  + 2 \xi \sum_{j\ell (j\neq\ell)} {\rm Im}[\mathsf{G}^{(j\ell)}_{dd}] \langle \hat{\sigma}_{j}^+ \hat{\sigma}_{\ell}^-\rangle.
\eeq
The coherent scattering rate \EQREF{eq:cohscatrateformula} is obtained by factorizing all terms $ \langle \hat{\sigma}_{j}^+ \hat{\sigma}_{\ell}^-\rangle\rightarrow  \langle \hat{\sigma}_{j}^+\>\< \hat{\sigma}_{\ell}^-\rangle$. This is straightforward to calculate for the phase-uniform mode and yields \EQREF{eq:nratecoh}, since $\langle \hat{\sigma}_{j}^\pm\>$ can be factored out of the sum. The incoherently scattered light includes both single-atom and many-atom contributions. Single-atom contributions are derived from $ \langle \hat{\sigma}_{j}^+ \hat{\sigma}_{j}^-\rangle -  \langle \hat{\sigma}_{j}^+\>\< \hat{\sigma}_{j}^-\rangle$. For the phase-uniform state, these yield 
\beq \label{eq:nrateincoh1}
n^{(1)}_{\rm inc}=2N\gamma (\rho_{ee}-|\rho_{ge}|^2). 
\eeq
On the other hand, the many-atom contributions are given by
\beq \label{eq:nrateincoh2}
n^{(2)}_{\rm inc}=2N (\gamma+\tilde\gamma) \big(\langle \hat{\sigma}_{j}^+ \hat{\sigma}_{\ell}^-\rangle - \langle \hat{\sigma}_{j}^+\>\< \hat{\sigma}_{\ell}^-\rangle \big),\quad j\neq\ell.
\eeq
The leading order $N$ dependence in the front of the expression is now $\propto N\tilde\gamma/\gamma$ [see Fig.~\ref{Fig:fluc}(a)], instead of the characteristic $\propto N$ or $\propto N^2$.
For fixed atomic positions, \EQREF{eq:nrateincoh2} signifies many-atom quantum effects. However, if atomic positions fluctuate, the corresponding expression can also contain significant classical contributions~\cite{Bettles2020,Ruostekoski2023}.


\begin{thebibliography}{69}%
\makeatletter
\providecommand \@ifxundefined [1]{%
 \@ifx{#1\undefined}
}%
\providecommand \@ifnum [1]{%
 \ifnum #1\expandafter \@firstoftwo
 \else \expandafter \@secondoftwo
 \fi
}%
\providecommand \@ifx [1]{%
 \ifx #1\expandafter \@firstoftwo
 \else \expandafter \@secondoftwo
 \fi
}%
\providecommand \natexlab [1]{#1}%
\providecommand \enquote  [1]{``#1''}%
\providecommand \bibnamefont  [1]{#1}%
\providecommand \bibfnamefont [1]{#1}%
\providecommand \citenamefont [1]{#1}%
\providecommand \href@noop [0]{\@secondoftwo}%
\providecommand \href [0]{\begingroup \@sanitize@url \@href}%
\providecommand \@href[1]{\@@startlink{#1}\@@href}%
\providecommand \@@href[1]{\endgroup#1\@@endlink}%
\providecommand \@sanitize@url [0]{\catcode `\\12\catcode `\$12\catcode
  `\&12\catcode `\#12\catcode `\^12\catcode `\_12\catcode `\%12\relax}%
\providecommand \@@startlink[1]{}%
\providecommand \@@endlink[0]{}%
\providecommand \url  [0]{\begingroup\@sanitize@url \@url }%
\providecommand \@url [1]{\endgroup\@href {#1}{\urlprefix }}%
\providecommand \urlprefix  [0]{URL }%
\providecommand \Eprint [0]{\href }%
\providecommand \doibase [0]{https://doi.org/}%
\providecommand \selectlanguage [0]{\@gobble}%
\providecommand \bibinfo  [0]{\@secondoftwo}%
\providecommand \bibfield  [0]{\@secondoftwo}%
\providecommand \translation [1]{[#1]}%
\providecommand \BibitemOpen [0]{}%
\providecommand \bibitemStop [0]{}%
\providecommand \bibitemNoStop [0]{.\EOS\space}%
\providecommand \EOS [0]{\spacefactor3000\relax}%
\providecommand \BibitemShut  [1]{\csname bibitem#1\endcsname}%
\let\auto@bib@innerbib\@empty
\bibitem [{\citenamefont {Bender}\ \emph {et~al.}(2010)\citenamefont {Bender},
  \citenamefont {Stehle}, \citenamefont {Slama}, \citenamefont {Kaiser},
  \citenamefont {Piovella}, \citenamefont {Zimmermann},\ and\ \citenamefont
  {Courteille}}]{Bender2010}%
  \BibitemOpen
  \bibfield  {author} {\bibinfo {author} {\bibfnamefont {H.}~\bibnamefont
  {Bender}}, \bibinfo {author} {\bibfnamefont {C.}~\bibnamefont {Stehle}},
  \bibinfo {author} {\bibfnamefont {S.}~\bibnamefont {Slama}}, \bibinfo
  {author} {\bibfnamefont {R.}~\bibnamefont {Kaiser}}, \bibinfo {author}
  {\bibfnamefont {N.}~\bibnamefont {Piovella}}, \bibinfo {author}
  {\bibfnamefont {C.}~\bibnamefont {Zimmermann}},\ and\ \bibinfo {author}
  {\bibfnamefont {P.~W.}\ \bibnamefont {Courteille}},\ }\bibfield  {title}
  {\bibinfo {title} {Observation of cooperative mie scattering from an
  ultracold atomic cloud},\ }\href {https://doi.org/10.1103/PhysRevA.82.011404}
  {\bibfield  {journal} {\bibinfo  {journal} {Phys. Rev. A}\ }\textbf {\bibinfo
  {volume} {82}},\ \bibinfo {pages} {011404(R)} (\bibinfo {year}
  {2010})}\BibitemShut {NoStop}%
\bibitem [{\citenamefont {Balik}\ \emph {et~al.}(2013)\citenamefont {Balik},
  \citenamefont {Win}, \citenamefont {Havey}, \citenamefont {Sokolov},\ and\
  \citenamefont {Kupriyanov}}]{BalikEtAl2013}%
  \BibitemOpen
  \bibfield  {author} {\bibinfo {author} {\bibfnamefont {S.}~\bibnamefont
  {Balik}}, \bibinfo {author} {\bibfnamefont {A.~L.}\ \bibnamefont {Win}},
  \bibinfo {author} {\bibfnamefont {M.~D.}\ \bibnamefont {Havey}}, \bibinfo
  {author} {\bibfnamefont {I.~M.}\ \bibnamefont {Sokolov}},\ and\ \bibinfo
  {author} {\bibfnamefont {D.~V.}\ \bibnamefont {Kupriyanov}},\ }\bibfield
  {title} {\bibinfo {title} {Near-resonance light scattering from a
  high-density ultracold atomic ${}^{87}${Rb} gas},\ }\href
  {https://doi.org/10.1103/PhysRevA.87.053817} {\bibfield  {journal} {\bibinfo
  {journal} {Phys. Rev. A}\ }\textbf {\bibinfo {volume} {87}},\ \bibinfo
  {pages} {053817} (\bibinfo {year} {2013})}\BibitemShut {NoStop}%
\bibitem [{\citenamefont {Pellegrino}\ \emph {et~al.}(2014)\citenamefont
  {Pellegrino}, \citenamefont {Bourgain}, \citenamefont {Jennewein},
  \citenamefont {Sortais}, \citenamefont {Browaeys}, \citenamefont {Jenkins},\
  and\ \citenamefont {Ruostekoski}}]{Pellegrino2014a}%
  \BibitemOpen
  \bibfield  {author} {\bibinfo {author} {\bibfnamefont {J.}~\bibnamefont
  {Pellegrino}}, \bibinfo {author} {\bibfnamefont {R.}~\bibnamefont
  {Bourgain}}, \bibinfo {author} {\bibfnamefont {S.}~\bibnamefont {Jennewein}},
  \bibinfo {author} {\bibfnamefont {Y.~R.~P.}\ \bibnamefont {Sortais}},
  \bibinfo {author} {\bibfnamefont {A.}~\bibnamefont {Browaeys}}, \bibinfo
  {author} {\bibfnamefont {S.~D.}\ \bibnamefont {Jenkins}},\ and\ \bibinfo
  {author} {\bibfnamefont {J.}~\bibnamefont {Ruostekoski}},\ }\bibfield
  {title} {\bibinfo {title} {Observation of suppression of light scattering
  induced by dipole-dipole interactions in a cold-atom ensemble},\ }\href
  {https://doi.org/10.1103/PhysRevLett.113.133602} {\bibfield  {journal}
  {\bibinfo  {journal} {Phys. Rev. Lett.}\ }\textbf {\bibinfo {volume} {113}},\
  \bibinfo {pages} {133602} (\bibinfo {year} {2014})}\BibitemShut {NoStop}%
\bibitem [{\citenamefont {Jenkins}\ \emph
  {et~al.}(2016{\natexlab{a}})\citenamefont {Jenkins}, \citenamefont
  {Ruostekoski}, \citenamefont {Javanainen}, \citenamefont {Bourgain},
  \citenamefont {Jennewein}, \citenamefont {Sortais},\ and\ \citenamefont
  {Browaeys}}]{Jenkins_thermshift}%
  \BibitemOpen
  \bibfield  {author} {\bibinfo {author} {\bibfnamefont {S.~D.}\ \bibnamefont
  {Jenkins}}, \bibinfo {author} {\bibfnamefont {J.}~\bibnamefont
  {Ruostekoski}}, \bibinfo {author} {\bibfnamefont {J.}~\bibnamefont
  {Javanainen}}, \bibinfo {author} {\bibfnamefont {R.}~\bibnamefont
  {Bourgain}}, \bibinfo {author} {\bibfnamefont {S.}~\bibnamefont {Jennewein}},
  \bibinfo {author} {\bibfnamefont {Y.~R.~P.}\ \bibnamefont {Sortais}},\ and\
  \bibinfo {author} {\bibfnamefont {A.}~\bibnamefont {Browaeys}},\ }\bibfield
  {title} {\bibinfo {title} {Optical resonance shifts in the fluorescence of
  thermal and cold atomic gases},\ }\href
  {https://doi.org/10.1103/PhysRevLett.116.183601} {\bibfield  {journal}
  {\bibinfo  {journal} {Phys. Rev. Lett.}\ }\textbf {\bibinfo {volume} {116}},\
  \bibinfo {pages} {183601} (\bibinfo {year} {2016}{\natexlab{a}})}\BibitemShut
  {NoStop}%
\bibitem [{\citenamefont {Jennewein}\ \emph {et~al.}(2016)\citenamefont
  {Jennewein}, \citenamefont {Besbes}, \citenamefont {Schilder}, \citenamefont
  {Jenkins}, \citenamefont {Sauvan}, \citenamefont {Ruostekoski}, \citenamefont
  {Greffet}, \citenamefont {Sortais},\ and\ \citenamefont
  {Browaeys}}]{Jennewein_trans}%
  \BibitemOpen
  \bibfield  {author} {\bibinfo {author} {\bibfnamefont {S.}~\bibnamefont
  {Jennewein}}, \bibinfo {author} {\bibfnamefont {M.}~\bibnamefont {Besbes}},
  \bibinfo {author} {\bibfnamefont {N.~J.}\ \bibnamefont {Schilder}}, \bibinfo
  {author} {\bibfnamefont {S.~D.}\ \bibnamefont {Jenkins}}, \bibinfo {author}
  {\bibfnamefont {C.}~\bibnamefont {Sauvan}}, \bibinfo {author} {\bibfnamefont
  {J.}~\bibnamefont {Ruostekoski}}, \bibinfo {author} {\bibfnamefont {J.-J.}\
  \bibnamefont {Greffet}}, \bibinfo {author} {\bibfnamefont {Y.~R.~P.}\
  \bibnamefont {Sortais}},\ and\ \bibinfo {author} {\bibfnamefont
  {A.}~\bibnamefont {Browaeys}},\ }\bibfield  {title} {\bibinfo {title}
  {Coherent scattering of near-resonant light by a dense microscopic cold
  atomic cloud},\ }\href {https://doi.org/10.1103/PhysRevLett.116.233601}
  {\bibfield  {journal} {\bibinfo  {journal} {Phys. Rev. Lett.}\ }\textbf
  {\bibinfo {volume} {116}},\ \bibinfo {pages} {233601} (\bibinfo {year}
  {2016})}\BibitemShut {NoStop}%
\bibitem [{\citenamefont {Corman}\ \emph {et~al.}(2017)\citenamefont {Corman},
  \citenamefont {Ville}, \citenamefont {Saint-Jalm}, \citenamefont
  {Aidelsburger}, \citenamefont {Bienaim\'e}, \citenamefont {Nascimb\`ene},
  \citenamefont {Dalibard},\ and\ \citenamefont {Beugnon}}]{Dalibard_slab}%
  \BibitemOpen
  \bibfield  {author} {\bibinfo {author} {\bibfnamefont {L.}~\bibnamefont
  {Corman}}, \bibinfo {author} {\bibfnamefont {J.~L.}\ \bibnamefont {Ville}},
  \bibinfo {author} {\bibfnamefont {R.}~\bibnamefont {Saint-Jalm}}, \bibinfo
  {author} {\bibfnamefont {M.}~\bibnamefont {Aidelsburger}}, \bibinfo {author}
  {\bibfnamefont {T.}~\bibnamefont {Bienaim\'e}}, \bibinfo {author}
  {\bibfnamefont {S.}~\bibnamefont {Nascimb\`ene}}, \bibinfo {author}
  {\bibfnamefont {J.}~\bibnamefont {Dalibard}},\ and\ \bibinfo {author}
  {\bibfnamefont {J.}~\bibnamefont {Beugnon}},\ }\bibfield  {title} {\bibinfo
  {title} {Transmission of near-resonant light through a dense slab of cold
  atoms},\ }\href {https://doi.org/10.1103/PhysRevA.96.053629} {\bibfield
  {journal} {\bibinfo  {journal} {Phys. Rev. A}\ }\textbf {\bibinfo {volume}
  {96}},\ \bibinfo {pages} {053629} (\bibinfo {year} {2017})}\BibitemShut
  {NoStop}%
\bibitem [{\citenamefont {Saint-Jalm}\ \emph {et~al.}(2018)\citenamefont
  {Saint-Jalm}, \citenamefont {Aidelsburger}, \citenamefont {Ville},
  \citenamefont {Corman}, \citenamefont {Hadzibabic}, \citenamefont {Delande},
  \citenamefont {Nascimbene}, \citenamefont {Cherroret}, \citenamefont
  {Dalibard},\ and\ \citenamefont {Beugnon}}]{Saint-Jalm2018}%
  \BibitemOpen
  \bibfield  {author} {\bibinfo {author} {\bibfnamefont {R.}~\bibnamefont
  {Saint-Jalm}}, \bibinfo {author} {\bibfnamefont {M.}~\bibnamefont
  {Aidelsburger}}, \bibinfo {author} {\bibfnamefont {J.~L.}\ \bibnamefont
  {Ville}}, \bibinfo {author} {\bibfnamefont {L.}~\bibnamefont {Corman}},
  \bibinfo {author} {\bibfnamefont {Z.}~\bibnamefont {Hadzibabic}}, \bibinfo
  {author} {\bibfnamefont {D.}~\bibnamefont {Delande}}, \bibinfo {author}
  {\bibfnamefont {S.}~\bibnamefont {Nascimbene}}, \bibinfo {author}
  {\bibfnamefont {N.}~\bibnamefont {Cherroret}}, \bibinfo {author}
  {\bibfnamefont {J.}~\bibnamefont {Dalibard}},\ and\ \bibinfo {author}
  {\bibfnamefont {J.}~\bibnamefont {Beugnon}},\ }\bibfield  {title} {\bibinfo
  {title} {{Resonant-light diffusion in a disordered atomic layer}},\ }\href
  {https://doi.org/10.1103/PhysRevA.97.061801} {\bibfield  {journal} {\bibinfo
  {journal} {Phys. Rev. A}\ }\textbf {\bibinfo {volume} {97}},\ \bibinfo
  {pages} {061801} (\bibinfo {year} {2018})}\BibitemShut {NoStop}%
\bibitem [{\citenamefont {Rui}\ \emph {et~al.}(2020)\citenamefont {Rui},
  \citenamefont {Wei}, \citenamefont {Rubio-Abadal}, \citenamefont {Hollerith},
  \citenamefont {Zeiher}, \citenamefont {Stamper-Kurn}, \citenamefont {Gross},\
  and\ \citenamefont {Bloch}}]{Rui2020}%
  \BibitemOpen
  \bibfield  {author} {\bibinfo {author} {\bibfnamefont {J.}~\bibnamefont
  {Rui}}, \bibinfo {author} {\bibfnamefont {D.}~\bibnamefont {Wei}}, \bibinfo
  {author} {\bibfnamefont {A.}~\bibnamefont {Rubio-Abadal}}, \bibinfo {author}
  {\bibfnamefont {S.}~\bibnamefont {Hollerith}}, \bibinfo {author}
  {\bibfnamefont {J.}~\bibnamefont {Zeiher}}, \bibinfo {author} {\bibfnamefont
  {D.~M.}\ \bibnamefont {Stamper-Kurn}}, \bibinfo {author} {\bibfnamefont
  {C.}~\bibnamefont {Gross}},\ and\ \bibinfo {author} {\bibfnamefont
  {I.}~\bibnamefont {Bloch}},\ }\bibfield  {title} {\bibinfo {title} {{A
  subradiant optical mirror formed by a single structured atomic layer}},\
  }\href {https://doi.org/10.1038/s41586-020-2463-x} {\bibfield  {journal}
  {\bibinfo  {journal} {Nature}\ }\textbf {\bibinfo {volume} {583}},\ \bibinfo
  {pages} {369} (\bibinfo {year} {2020})}\BibitemShut {NoStop}%
\bibitem [{\citenamefont {Ferioli}\ \emph {et~al.}(2021)\citenamefont
  {Ferioli}, \citenamefont {Glicenstein}, \citenamefont {Henriet},
  \citenamefont {Ferrier-Barbut},\ and\ \citenamefont {Browaeys}}]{Ferioli21}%
  \BibitemOpen
  \bibfield  {author} {\bibinfo {author} {\bibfnamefont {G.}~\bibnamefont
  {Ferioli}}, \bibinfo {author} {\bibfnamefont {A.}~\bibnamefont
  {Glicenstein}}, \bibinfo {author} {\bibfnamefont {L.}~\bibnamefont
  {Henriet}}, \bibinfo {author} {\bibfnamefont {I.}~\bibnamefont
  {Ferrier-Barbut}},\ and\ \bibinfo {author} {\bibfnamefont {A.}~\bibnamefont
  {Browaeys}},\ }\bibfield  {title} {\bibinfo {title} {Storage and release of
  subradiant excitations in a dense atomic cloud},\ }\href
  {https://doi.org/10.1103/PhysRevX.11.021031} {\bibfield  {journal} {\bibinfo
  {journal} {Phys. Rev. X}\ }\textbf {\bibinfo {volume} {11}},\ \bibinfo
  {pages} {021031} (\bibinfo {year} {2021})}\BibitemShut {NoStop}%
\bibitem [{\citenamefont {Srakaew}\ \emph {et~al.}(2023)\citenamefont
  {Srakaew}, \citenamefont {Weckesser}, \citenamefont {Hollerith},
  \citenamefont {Wei}, \citenamefont {Adler}, \citenamefont {Bloch},\ and\
  \citenamefont {Zeiher}}]{Srakaew22}%
  \BibitemOpen
  \bibfield  {author} {\bibinfo {author} {\bibfnamefont {K.}~\bibnamefont
  {Srakaew}}, \bibinfo {author} {\bibfnamefont {P.}~\bibnamefont {Weckesser}},
  \bibinfo {author} {\bibfnamefont {S.}~\bibnamefont {Hollerith}}, \bibinfo
  {author} {\bibfnamefont {D.}~\bibnamefont {Wei}}, \bibinfo {author}
  {\bibfnamefont {D.}~\bibnamefont {Adler}}, \bibinfo {author} {\bibfnamefont
  {I.}~\bibnamefont {Bloch}},\ and\ \bibinfo {author} {\bibfnamefont
  {J.}~\bibnamefont {Zeiher}},\ }\bibfield  {title} {\bibinfo {title} {A
  subwavelength atomic array switched by a single rydberg atom},\ }\href
  {https://doi.org/10.1038/s41567-023-01959-y} {\bibfield  {journal} {\bibinfo
  {journal} {Nature Physics}\ }\textbf {\bibinfo {volume} {19}},\ \bibinfo
  {pages} {714} (\bibinfo {year} {2023})}\BibitemShut {NoStop}%
\bibitem [{\citenamefont {Dicke}(1954)}]{Dicke54}%
  \BibitemOpen
  \bibfield  {author} {\bibinfo {author} {\bibfnamefont {R.~H.}\ \bibnamefont
  {Dicke}},\ }\bibfield  {title} {\bibinfo {title} {Coherence in spontaneous
  radiation processes},\ }\href {https://doi.org/10.1103/PhysRev.93.99}
  {\bibfield  {journal} {\bibinfo  {journal} {Phys. Rev.}\ }\textbf {\bibinfo
  {volume} {93}},\ \bibinfo {pages} {99} (\bibinfo {year} {1954})}\BibitemShut
  {NoStop}%
\bibitem [{\citenamefont {Svidzinsky}\ \emph {et~al.}(2008)\citenamefont
  {Svidzinsky}, \citenamefont {Chang},\ and\ \citenamefont
  {Scully}}]{Svidzinsky08}%
  \BibitemOpen
  \bibfield  {author} {\bibinfo {author} {\bibfnamefont {A.~A.}\ \bibnamefont
  {Svidzinsky}}, \bibinfo {author} {\bibfnamefont {J.-T.}\ \bibnamefont
  {Chang}},\ and\ \bibinfo {author} {\bibfnamefont {M.~O.}\ \bibnamefont
  {Scully}},\ }\bibfield  {title} {\bibinfo {title} {Dynamical evolution of
  correlated spontaneous emission of a single photon from a uniformly excited
  cloud of $n$ atoms},\ }\href {https://doi.org/10.1103/PhysRevLett.100.160504}
  {\bibfield  {journal} {\bibinfo  {journal} {Phys. Rev. Lett.}\ }\textbf
  {\bibinfo {volume} {100}},\ \bibinfo {pages} {160504} (\bibinfo {year}
  {2008})}\BibitemShut {NoStop}%
\bibitem [{\citenamefont {Kwong}\ \emph {et~al.}(2014)\citenamefont {Kwong},
  \citenamefont {Yang}, \citenamefont {Pramod}, \citenamefont {Pandey},
  \citenamefont {Delande}, \citenamefont {Pierrat},\ and\ \citenamefont
  {Wilkowski}}]{wilkowski}%
  \BibitemOpen
  \bibfield  {author} {\bibinfo {author} {\bibfnamefont {C.~C.}\ \bibnamefont
  {Kwong}}, \bibinfo {author} {\bibfnamefont {T.}~\bibnamefont {Yang}},
  \bibinfo {author} {\bibfnamefont {M.~S.}\ \bibnamefont {Pramod}}, \bibinfo
  {author} {\bibfnamefont {K.}~\bibnamefont {Pandey}}, \bibinfo {author}
  {\bibfnamefont {D.}~\bibnamefont {Delande}}, \bibinfo {author} {\bibfnamefont
  {R.}~\bibnamefont {Pierrat}},\ and\ \bibinfo {author} {\bibfnamefont
  {D.}~\bibnamefont {Wilkowski}},\ }\bibfield  {title} {\bibinfo {title}
  {Cooperative emission of a coherent superflash of light},\ }\href
  {https://doi.org/10.1103/PhysRevLett.113.223601} {\bibfield  {journal}
  {\bibinfo  {journal} {Phys. Rev. Lett.}\ }\textbf {\bibinfo {volume} {113}},\
  \bibinfo {pages} {223601} (\bibinfo {year} {2014})}\BibitemShut {NoStop}%
\bibitem [{\citenamefont {Ara\'ujo}\ \emph {et~al.}(2016)\citenamefont
  {Ara\'ujo}, \citenamefont {Kre\ifmmode \check{s}\else
  \v{s}\fi{}i\ifmmode~\acute{c}\else \'{c}\fi{}}, \citenamefont {Kaiser},\ and\
  \citenamefont {Guerin}}]{Araujo16}%
  \BibitemOpen
  \bibfield  {author} {\bibinfo {author} {\bibfnamefont {M.~O.}\ \bibnamefont
  {Ara\'ujo}}, \bibinfo {author} {\bibfnamefont {I.}~\bibnamefont {Kre\ifmmode
  \check{s}\else \v{s}\fi{}i\ifmmode~\acute{c}\else \'{c}\fi{}}}, \bibinfo
  {author} {\bibfnamefont {R.}~\bibnamefont {Kaiser}},\ and\ \bibinfo {author}
  {\bibfnamefont {W.}~\bibnamefont {Guerin}},\ }\bibfield  {title} {\bibinfo
  {title} {Superradiance in a large and dilute cloud of cold atoms in the
  linear-optics regime},\ }\href
  {https://doi.org/10.1103/PhysRevLett.117.073002} {\bibfield  {journal}
  {\bibinfo  {journal} {Phys. Rev. Lett.}\ }\textbf {\bibinfo {volume} {117}},\
  \bibinfo {pages} {073002} (\bibinfo {year} {2016})}\BibitemShut {NoStop}%
\bibitem [{\citenamefont {Roof}\ \emph {et~al.}(2016)\citenamefont {Roof},
  \citenamefont {Kemp}, \citenamefont {Havey},\ and\ \citenamefont
  {Sokolov}}]{Roof16}%
  \BibitemOpen
  \bibfield  {author} {\bibinfo {author} {\bibfnamefont {S.~J.}\ \bibnamefont
  {Roof}}, \bibinfo {author} {\bibfnamefont {K.~J.}\ \bibnamefont {Kemp}},
  \bibinfo {author} {\bibfnamefont {M.~D.}\ \bibnamefont {Havey}},\ and\
  \bibinfo {author} {\bibfnamefont {I.~M.}\ \bibnamefont {Sokolov}},\
  }\bibfield  {title} {\bibinfo {title} {Observation of single-photon
  superradiance and the cooperative lamb shift in an extended sample of cold
  atoms},\ }\href {https://doi.org/10.1103/PhysRevLett.117.073003} {\bibfield
  {journal} {\bibinfo  {journal} {Phys. Rev. Lett.}\ }\textbf {\bibinfo
  {volume} {117}},\ \bibinfo {pages} {073003} (\bibinfo {year}
  {2016})}\BibitemShut {NoStop}%
\bibitem [{\citenamefont {Solano}\ \emph {et~al.}(2017)\citenamefont {Solano},
  \citenamefont {Barberis-Blostein}, \citenamefont {Fatemi}, \citenamefont
  {Orozco},\ and\ \citenamefont {Rolston}}]{Solano_super}%
  \BibitemOpen
  \bibfield  {author} {\bibinfo {author} {\bibfnamefont {P.}~\bibnamefont
  {Solano}}, \bibinfo {author} {\bibfnamefont {P.}~\bibnamefont
  {Barberis-Blostein}}, \bibinfo {author} {\bibfnamefont {F.~K.}\ \bibnamefont
  {Fatemi}}, \bibinfo {author} {\bibfnamefont {L.~A.}\ \bibnamefont {Orozco}},\
  and\ \bibinfo {author} {\bibfnamefont {S.~L.}\ \bibnamefont {Rolston}},\
  }\bibfield  {title} {\bibinfo {title} {Super-radiance reveals infinite-range
  dipole interactions through a nanofiber},\ }\href
  {https://doi.org/10.1038/s41467-017-01994-3} {\bibfield  {journal} {\bibinfo
  {journal} {Nature Communications}\ }\textbf {\bibinfo {volume} {8}},\
  \bibinfo {pages} {1857} (\bibinfo {year} {2017})}\BibitemShut {NoStop}%
\bibitem [{\citenamefont {Sutherland}\ and\ \citenamefont
  {Robicheaux}(2017)}]{Sutherland2017}%
  \BibitemOpen
  \bibfield  {author} {\bibinfo {author} {\bibfnamefont {R.~T.}\ \bibnamefont
  {Sutherland}}\ and\ \bibinfo {author} {\bibfnamefont {F.}~\bibnamefont
  {Robicheaux}},\ }\bibfield  {title} {\bibinfo {title} {Superradiance in
  inverted multilevel atomic clouds},\ }\href
  {https://doi.org/10.1103/PhysRevA.95.033839} {\bibfield  {journal} {\bibinfo
  {journal} {Phys. Rev. A}\ }\textbf {\bibinfo {volume} {95}},\ \bibinfo
  {pages} {033839} (\bibinfo {year} {2017})}\BibitemShut {NoStop}%
\bibitem [{\citenamefont {Okaba}\ \emph {et~al.}(2019)\citenamefont {Okaba},
  \citenamefont {Yu}, \citenamefont {Vincetti}, \citenamefont {Benabid},\ and\
  \citenamefont {Katori}}]{Okaba2019}%
  \BibitemOpen
  \bibfield  {author} {\bibinfo {author} {\bibfnamefont {S.}~\bibnamefont
  {Okaba}}, \bibinfo {author} {\bibfnamefont {D.}~\bibnamefont {Yu}}, \bibinfo
  {author} {\bibfnamefont {L.}~\bibnamefont {Vincetti}}, \bibinfo {author}
  {\bibfnamefont {F.}~\bibnamefont {Benabid}},\ and\ \bibinfo {author}
  {\bibfnamefont {H.}~\bibnamefont {Katori}},\ }\bibfield  {title} {\bibinfo
  {title} {Superradiance from lattice-confined atoms inside hollow core
  fibre},\ }\href {https://doi.org/10.1038/s42005-019-0237-2} {\bibfield
  {journal} {\bibinfo  {journal} {Communications Physics}\ }\textbf {\bibinfo
  {volume} {2}},\ \bibinfo {pages} {136} (\bibinfo {year} {2019})}\BibitemShut
  {NoStop}%
\bibitem [{\citenamefont {Robicheaux}(2021)}]{Robicheaux2021}%
  \BibitemOpen
  \bibfield  {author} {\bibinfo {author} {\bibfnamefont {F.}~\bibnamefont
  {Robicheaux}},\ }\bibfield  {title} {\bibinfo {title} {Theoretical study of
  early-time superradiance for atom clouds and arrays},\ }\href
  {https://doi.org/10.1103/PhysRevA.104.063706} {\bibfield  {journal} {\bibinfo
   {journal} {Phys. Rev. A}\ }\textbf {\bibinfo {volume} {104}},\ \bibinfo
  {pages} {063706} (\bibinfo {year} {2021})}\BibitemShut {NoStop}%
\bibitem [{\citenamefont {Pi\~neiro Orioli}\ \emph {et~al.}(2022)\citenamefont
  {Pi\~neiro Orioli}, \citenamefont {Thompson},\ and\ \citenamefont
  {Rey}}]{Orioli2022}%
  \BibitemOpen
  \bibfield  {author} {\bibinfo {author} {\bibfnamefont {A.}~\bibnamefont
  {Pi\~neiro Orioli}}, \bibinfo {author} {\bibfnamefont {J.~K.}\ \bibnamefont
  {Thompson}},\ and\ \bibinfo {author} {\bibfnamefont {A.~M.}\ \bibnamefont
  {Rey}},\ }\bibfield  {title} {\bibinfo {title} {Emergent dark states from
  superradiant dynamics in multilevel atoms in a cavity},\ }\href
  {https://doi.org/10.1103/PhysRevX.12.011054} {\bibfield  {journal} {\bibinfo
  {journal} {Phys. Rev. X}\ }\textbf {\bibinfo {volume} {12}},\ \bibinfo
  {pages} {011054} (\bibinfo {year} {2022})}\BibitemShut {NoStop}%
\bibitem [{\citenamefont {Masson}\ and\ \citenamefont
  {Asenjo-Garcia}(2022)}]{Masson2022}%
  \BibitemOpen
  \bibfield  {author} {\bibinfo {author} {\bibfnamefont {S.~J.}\ \bibnamefont
  {Masson}}\ and\ \bibinfo {author} {\bibfnamefont {A.}~\bibnamefont
  {Asenjo-Garcia}},\ }\bibfield  {title} {\bibinfo {title} {Universality of
  dicke superradiance in arrays of quantum emitters},\ }\href
  {https://doi.org/10.1038/s41467-022-29805-4} {\bibfield  {journal} {\bibinfo
  {journal} {Nature Communications}\ }\textbf {\bibinfo {volume} {13}},\
  \bibinfo {pages} {2285} (\bibinfo {year} {2022})}\BibitemShut {NoStop}%
\bibitem [{\citenamefont {Asselie}\ \emph {et~al.}(2022)\citenamefont
  {Asselie}, \citenamefont {Cipris},\ and\ \citenamefont
  {Guerin}}]{Asselie2022}%
  \BibitemOpen
  \bibfield  {author} {\bibinfo {author} {\bibfnamefont {S.}~\bibnamefont
  {Asselie}}, \bibinfo {author} {\bibfnamefont {A.}~\bibnamefont {Cipris}},\
  and\ \bibinfo {author} {\bibfnamefont {W.}~\bibnamefont {Guerin}},\
  }\bibfield  {title} {\bibinfo {title} {Optical interpretation of
  linear-optics superradiance and subradiance},\ }\href
  {https://doi.org/10.1103/PhysRevA.106.063712} {\bibfield  {journal} {\bibinfo
   {journal} {Phys. Rev. A}\ }\textbf {\bibinfo {volume} {106}},\ \bibinfo
  {pages} {063712} (\bibinfo {year} {2022})}\BibitemShut {NoStop}%
\bibitem [{\citenamefont {Malz}\ \emph {et~al.}(2022)\citenamefont {Malz},
  \citenamefont {Trivedi},\ and\ \citenamefont {Cirac}}]{Malz2022}%
  \BibitemOpen
  \bibfield  {author} {\bibinfo {author} {\bibfnamefont {D.}~\bibnamefont
  {Malz}}, \bibinfo {author} {\bibfnamefont {R.}~\bibnamefont {Trivedi}},\ and\
  \bibinfo {author} {\bibfnamefont {J.~I.}\ \bibnamefont {Cirac}},\ }\bibfield
  {title} {\bibinfo {title} {Large-$n$ limit of dicke superradiance},\ }\href
  {https://doi.org/10.1103/PhysRevA.106.013716} {\bibfield  {journal} {\bibinfo
   {journal} {Phys. Rev. A}\ }\textbf {\bibinfo {volume} {106}},\ \bibinfo
  {pages} {013716} (\bibinfo {year} {2022})}\BibitemShut {NoStop}%
\bibitem [{\citenamefont {Rubies-Bigorda}\ \emph {et~al.}(2023)\citenamefont
  {Rubies-Bigorda}, \citenamefont {Ostermann},\ and\ \citenamefont
  {Yelin}}]{Rubies-Bigorda2023}%
  \BibitemOpen
  \bibfield  {author} {\bibinfo {author} {\bibfnamefont {O.}~\bibnamefont
  {Rubies-Bigorda}}, \bibinfo {author} {\bibfnamefont {S.}~\bibnamefont
  {Ostermann}},\ and\ \bibinfo {author} {\bibfnamefont {S.~F.}\ \bibnamefont
  {Yelin}},\ }\bibfield  {title} {\bibinfo {title} {Characterizing superradiant
  dynamics in atomic arrays via a cumulant expansion approach},\ }\href
  {https://doi.org/10.1103/PhysRevResearch.5.013091} {\bibfield  {journal}
  {\bibinfo  {journal} {Phys. Rev. Res.}\ }\textbf {\bibinfo {volume} {5}},\
  \bibinfo {pages} {013091} (\bibinfo {year} {2023})}\BibitemShut {NoStop}%
\bibitem [{\citenamefont {Liedl}\ \emph {et~al.}(2024)\citenamefont {Liedl},
  \citenamefont {Tebbenjohanns}, \citenamefont {Bach}, \citenamefont {Pucher},
  \citenamefont {Rauschenbeutel},\ and\ \citenamefont
  {Schneeweiss}}]{Liedl2024}%
  \BibitemOpen
  \bibfield  {author} {\bibinfo {author} {\bibfnamefont {C.}~\bibnamefont
  {Liedl}}, \bibinfo {author} {\bibfnamefont {F.}~\bibnamefont
  {Tebbenjohanns}}, \bibinfo {author} {\bibfnamefont {C.}~\bibnamefont {Bach}},
  \bibinfo {author} {\bibfnamefont {S.}~\bibnamefont {Pucher}}, \bibinfo
  {author} {\bibfnamefont {A.}~\bibnamefont {Rauschenbeutel}},\ and\ \bibinfo
  {author} {\bibfnamefont {P.}~\bibnamefont {Schneeweiss}},\ }\bibfield
  {title} {\bibinfo {title} {Observation of superradiant bursts in a cascaded
  quantum system},\ }\href {https://doi.org/10.1103/PhysRevX.14.011020}
  {\bibfield  {journal} {\bibinfo  {journal} {Phys. Rev. X}\ }\textbf {\bibinfo
  {volume} {14}},\ \bibinfo {pages} {011020} (\bibinfo {year}
  {2024})}\BibitemShut {NoStop}%
\bibitem [{\citenamefont {Senitzky}(1972)}]{Senitzky1972}%
  \BibitemOpen
  \bibfield  {author} {\bibinfo {author} {\bibfnamefont {I.~R.}\ \bibnamefont
  {Senitzky}},\ }\bibfield  {title} {\bibinfo {title} {Interaction between a
  nonlinear oscillator and a radiation field},\ }\href
  {https://doi.org/10.1103/PhysRevA.6.1175} {\bibfield  {journal} {\bibinfo
  {journal} {Phys. Rev. A}\ }\textbf {\bibinfo {volume} {6}},\ \bibinfo {pages}
  {1175} (\bibinfo {year} {1972})}\BibitemShut {NoStop}%
\bibitem [{\citenamefont {Agarwal}\ \emph {et~al.}(1977)\citenamefont
  {Agarwal}, \citenamefont {Brown}, \citenamefont {Narducci},\ and\
  \citenamefont {Vetri}}]{Agarwal1977}%
  \BibitemOpen
  \bibfield  {author} {\bibinfo {author} {\bibfnamefont {G.~S.}\ \bibnamefont
  {Agarwal}}, \bibinfo {author} {\bibfnamefont {A.~C.}\ \bibnamefont {Brown}},
  \bibinfo {author} {\bibfnamefont {L.~M.}\ \bibnamefont {Narducci}},\ and\
  \bibinfo {author} {\bibfnamefont {G.}~\bibnamefont {Vetri}},\ }\bibfield
  {title} {\bibinfo {title} {Collective atomic effects in resonance
  fluorescence},\ }\href {https://doi.org/10.1103/PhysRevA.15.1613} {\bibfield
  {journal} {\bibinfo  {journal} {Phys. Rev. A}\ }\textbf {\bibinfo {volume}
  {15}},\ \bibinfo {pages} {1613} (\bibinfo {year} {1977})}\BibitemShut
  {NoStop}%
\bibitem [{\citenamefont {Drummond}\ and\ \citenamefont
  {Carmichael}(1978)}]{Drummond1978}%
  \BibitemOpen
  \bibfield  {author} {\bibinfo {author} {\bibfnamefont {P.}~\bibnamefont
  {Drummond}}\ and\ \bibinfo {author} {\bibfnamefont {H.}~\bibnamefont
  {Carmichael}},\ }\bibfield  {title} {\bibinfo {title} {Volterra cycles and
  the cooperative fluorescence critical point},\ }\href
  {https://doi.org/https://doi.org/10.1016/0030-4018(78)90198-0} {\bibfield
  {journal} {\bibinfo  {journal} {Optics Communications}\ }\textbf {\bibinfo
  {volume} {27}},\ \bibinfo {pages} {160} (\bibinfo {year} {1978})}\BibitemShut
  {NoStop}%
\bibitem [{\citenamefont {Walls}\ \emph {et~al.}(1978)\citenamefont {Walls},
  \citenamefont {Drummond}, \citenamefont {Hassan},\ and\ \citenamefont
  {Carmichael}}]{Walls1978}%
  \BibitemOpen
  \bibfield  {author} {\bibinfo {author} {\bibfnamefont {D.~F.}\ \bibnamefont
  {Walls}}, \bibinfo {author} {\bibfnamefont {P.~D.}\ \bibnamefont {Drummond}},
  \bibinfo {author} {\bibfnamefont {S.~S.}\ \bibnamefont {Hassan}},\ and\
  \bibinfo {author} {\bibfnamefont {H.~J.}\ \bibnamefont {Carmichael}},\
  }\bibfield  {title} {\bibinfo {title} {{Non-Equilibrium Phase Transitions in
  Cooperative Atomic Systems}},\ }\href {https://doi.org/10.1143/PTPS.64.307}
  {\bibfield  {journal} {\bibinfo  {journal} {Progress of Theoretical Physics
  Supplement}\ }\textbf {\bibinfo {volume} {64}},\ \bibinfo {pages} {307}
  (\bibinfo {year} {1978})}\BibitemShut {NoStop}%
\bibitem [{\citenamefont {Narducci}\ \emph {et~al.}(1978)\citenamefont
  {Narducci}, \citenamefont {Feng}, \citenamefont {Gilmore},\ and\
  \citenamefont {Agarwal}}]{Narducci1978}%
  \BibitemOpen
  \bibfield  {author} {\bibinfo {author} {\bibfnamefont {L.~M.}\ \bibnamefont
  {Narducci}}, \bibinfo {author} {\bibfnamefont {D.~H.}\ \bibnamefont {Feng}},
  \bibinfo {author} {\bibfnamefont {R.}~\bibnamefont {Gilmore}},\ and\ \bibinfo
  {author} {\bibfnamefont {G.~S.}\ \bibnamefont {Agarwal}},\ }\bibfield
  {title} {\bibinfo {title} {Transient and steady-state behavior of collective
  atomic systems driven by a classical field},\ }\href
  {https://doi.org/10.1103/PhysRevA.18.1571} {\bibfield  {journal} {\bibinfo
  {journal} {Phys. Rev. A}\ }\textbf {\bibinfo {volume} {18}},\ \bibinfo
  {pages} {1571} (\bibinfo {year} {1978})}\BibitemShut {NoStop}%
\bibitem [{\citenamefont {Puri}\ and\ \citenamefont
  {Lawande}(1979)}]{Puri1979}%
  \BibitemOpen
  \bibfield  {author} {\bibinfo {author} {\bibfnamefont {R.}~\bibnamefont
  {Puri}}\ and\ \bibinfo {author} {\bibfnamefont {S.}~\bibnamefont {Lawande}},\
  }\bibfield  {title} {\bibinfo {title} {Exact steady-state density operator
  for a collective atomic system in an external field},\ }\href
  {https://doi.org/https://doi.org/10.1016/0375-9601(79)90003-3} {\bibfield
  {journal} {\bibinfo  {journal} {Physics Letters A}\ }\textbf {\bibinfo
  {volume} {72}},\ \bibinfo {pages} {200} (\bibinfo {year} {1979})}\BibitemShut
  {NoStop}%
\bibitem [{\citenamefont {Carmichael}(1980)}]{Carmichael1980}%
  \BibitemOpen
  \bibfield  {author} {\bibinfo {author} {\bibfnamefont {H.~J.}\ \bibnamefont
  {Carmichael}},\ }\bibfield  {title} {\bibinfo {title} {Analytical and
  numerical results for the steady state in cooperative resonance
  fluorescence},\ }\href {https://doi.org/10.1088/0022-3700/13/18/009}
  {\bibfield  {journal} {\bibinfo  {journal} {Journal of Physics B: Atomic and
  Molecular Physics}\ }\textbf {\bibinfo {volume} {13}},\ \bibinfo {pages}
  {3551} (\bibinfo {year} {1980})}\BibitemShut {NoStop}%
\bibitem [{\citenamefont {Drummond}(1980)}]{Drummond1980b}%
  \BibitemOpen
  \bibfield  {author} {\bibinfo {author} {\bibfnamefont {P.~D.}\ \bibnamefont
  {Drummond}},\ }\bibfield  {title} {\bibinfo {title} {Observables and moments
  of cooperative resonance fluorescence},\ }\href
  {https://doi.org/10.1103/PhysRevA.22.1179} {\bibfield  {journal} {\bibinfo
  {journal} {Phys. Rev. A}\ }\textbf {\bibinfo {volume} {22}},\ \bibinfo
  {pages} {1179} (\bibinfo {year} {1980})}\BibitemShut {NoStop}%
\bibitem [{\citenamefont {Ferioli}\ \emph {et~al.}(2023)\citenamefont
  {Ferioli}, \citenamefont {Glicenstein}, \citenamefont {Ferrier-Barbut},\ and\
  \citenamefont {Browaeys}}]{Ferioli2023}%
  \BibitemOpen
  \bibfield  {author} {\bibinfo {author} {\bibfnamefont {G.}~\bibnamefont
  {Ferioli}}, \bibinfo {author} {\bibfnamefont {A.}~\bibnamefont
  {Glicenstein}}, \bibinfo {author} {\bibfnamefont {I.}~\bibnamefont
  {Ferrier-Barbut}},\ and\ \bibinfo {author} {\bibfnamefont {A.}~\bibnamefont
  {Browaeys}},\ }\bibfield  {title} {\bibinfo {title} {A non-equilibrium
  superradiant phase transition in free space},\ }\href
  {https://doi.org/10.1038/s41567-023-02064-w} {\bibfield  {journal} {\bibinfo
  {journal} {Nature Physics}\ }\textbf {\bibinfo {volume} {19}},\ \bibinfo
  {pages} {1345} (\bibinfo {year} {2023})}\BibitemShut {NoStop}%
\bibitem [{\citenamefont {Bonifacio}\ and\ \citenamefont
  {Lugiato}(1976)}]{bonifacio1976}%
  \BibitemOpen
  \bibfield  {author} {\bibinfo {author} {\bibfnamefont {R.}~\bibnamefont
  {Bonifacio}}\ and\ \bibinfo {author} {\bibfnamefont {L.}~\bibnamefont
  {Lugiato}},\ }\bibfield  {title} {\bibinfo {title} {Cooperative effects and
  bistability for resonance fluorescence},\ }\href
  {https://doi.org/https://doi.org/10.1016/0030-4018(76)90335-7} {\bibfield
  {journal} {\bibinfo  {journal} {Optics Communications}\ }\textbf {\bibinfo
  {volume} {19}},\ \bibinfo {pages} {172 } (\bibinfo {year}
  {1976})}\BibitemShut {NoStop}%
\bibitem [{\citenamefont {Bonifacio}\ and\ \citenamefont
  {Lugiato}(1978)}]{Bonifacio1978}%
  \BibitemOpen
  \bibfield  {author} {\bibinfo {author} {\bibfnamefont {R.}~\bibnamefont
  {Bonifacio}}\ and\ \bibinfo {author} {\bibfnamefont {L.~A.}\ \bibnamefont
  {Lugiato}},\ }\bibfield  {title} {\bibinfo {title} {{Optical bistability and
  cooperative effects in resonance fluorescence}},\ }\href
  {https://doi.org/10.1103/PhysRevA.18.1129} {\bibfield  {journal} {\bibinfo
  {journal} {Phys. Rev. A}\ }\textbf {\bibinfo {volume} {18}},\ \bibinfo
  {pages} {1129} (\bibinfo {year} {1978})}\BibitemShut {NoStop}%
\bibitem [{\citenamefont {Carmichael}\ and\ \citenamefont
  {Walls}(1977)}]{Carmichael1977}%
  \BibitemOpen
  \bibfield  {author} {\bibinfo {author} {\bibfnamefont {H.~J.}\ \bibnamefont
  {Carmichael}}\ and\ \bibinfo {author} {\bibfnamefont {D.~F.}\ \bibnamefont
  {Walls}},\ }\bibfield  {title} {\bibinfo {title} {Hysteresis in the spectrum
  for cooperative resonance fluorescence},\ }\href
  {https://doi.org/10.1088/0022-3700/10/18/002} {\bibfield  {journal} {\bibinfo
   {journal} {Journal of Physics B: Atomic and Molecular Physics}\ }\textbf
  {\bibinfo {volume} {10}},\ \bibinfo {pages} {L685} (\bibinfo {year}
  {1977})}\BibitemShut {NoStop}%
\bibitem [{\citenamefont {Agrawal}\ and\ \citenamefont
  {Carmichael}(1979)}]{Agrawal79}%
  \BibitemOpen
  \bibfield  {author} {\bibinfo {author} {\bibfnamefont {G.~P.}\ \bibnamefont
  {Agrawal}}\ and\ \bibinfo {author} {\bibfnamefont {H.~J.}\ \bibnamefont
  {Carmichael}},\ }\bibfield  {title} {\bibinfo {title} {{Optical bistability
  through nonlinear dispersion and absorption}},\ }\href
  {https://doi.org/10.1103/PhysRevA.19.2074} {\bibfield  {journal} {\bibinfo
  {journal} {Phys. Rev. A}\ }\textbf {\bibinfo {volume} {19}},\ \bibinfo
  {pages} {2074} (\bibinfo {year} {1979})}\BibitemShut {NoStop}%
\bibitem [{\citenamefont {Drummond}\ and\ \citenamefont
  {Walls}(1980)}]{Drummond1980}%
  \BibitemOpen
  \bibfield  {author} {\bibinfo {author} {\bibfnamefont {P.~D.}\ \bibnamefont
  {Drummond}}\ and\ \bibinfo {author} {\bibfnamefont {D.~F.}\ \bibnamefont
  {Walls}},\ }\bibfield  {title} {\bibinfo {title} {{Quantum theory of optical
  bistability. I. Nonlinear polarisability model}},\ }\href
  {https://doi.org/10.1088/0305-4470/13/2/034} {\bibfield  {journal} {\bibinfo
  {journal} {Journal of Physics A: Mathematical and General}\ }\textbf
  {\bibinfo {volume} {13}},\ \bibinfo {pages} {725} (\bibinfo {year}
  {1980})}\BibitemShut {NoStop}%
\bibitem [{\citenamefont {Drummond}\ and\ \citenamefont
  {Walls}(1981)}]{Drummond1981}%
  \BibitemOpen
  \bibfield  {author} {\bibinfo {author} {\bibfnamefont {P.~D.}\ \bibnamefont
  {Drummond}}\ and\ \bibinfo {author} {\bibfnamefont {D.~F.}\ \bibnamefont
  {Walls}},\ }\bibfield  {title} {\bibinfo {title} {{Quantum theory of optical
  bistability. II. Atomic fluorescence in a high-$Q$ cavity}},\ }\href
  {https://doi.org/10.1103/PhysRevA.23.2563} {\bibfield  {journal} {\bibinfo
  {journal} {Phys. Rev. A}\ }\textbf {\bibinfo {volume} {23}},\ \bibinfo
  {pages} {2563} (\bibinfo {year} {1981})}\BibitemShut {NoStop}%
\bibitem [{\citenamefont {Parmee}\ and\ \citenamefont
  {Ruostekoski}(2021)}]{Parmee2021}%
  \BibitemOpen
  \bibfield  {author} {\bibinfo {author} {\bibfnamefont {C.~D.}\ \bibnamefont
  {Parmee}}\ and\ \bibinfo {author} {\bibfnamefont {J.}~\bibnamefont
  {Ruostekoski}},\ }\bibfield  {title} {\bibinfo {title} {Bistable optical
  transmission through arrays of atoms in free space},\ }\href
  {https://doi.org/10.1103/PhysRevA.103.033706} {\bibfield  {journal} {\bibinfo
   {journal} {Phys. Rev. A}\ }\textbf {\bibinfo {volume} {103}},\ \bibinfo
  {pages} {033706} (\bibinfo {year} {2021})}\BibitemShut {NoStop}%
\bibitem [{\citenamefont {Agarwal}\ \emph {et~al.}(2024)\citenamefont
  {Agarwal}, \citenamefont {Chaparro}, \citenamefont {Barberena}, \citenamefont
  {Orioli}, \citenamefont {Ferioli}, \citenamefont {Pancaldi}, \citenamefont
  {Ferrier-Barbut}, \citenamefont {Browaeys},\ and\ \citenamefont
  {Rey}}]{agarwal2024}%
  \BibitemOpen
  \bibfield  {author} {\bibinfo {author} {\bibfnamefont {S.}~\bibnamefont
  {Agarwal}}, \bibinfo {author} {\bibfnamefont {E.}~\bibnamefont {Chaparro}},
  \bibinfo {author} {\bibfnamefont {D.}~\bibnamefont {Barberena}}, \bibinfo
  {author} {\bibfnamefont {A.~P.}\ \bibnamefont {Orioli}}, \bibinfo {author}
  {\bibfnamefont {G.}~\bibnamefont {Ferioli}}, \bibinfo {author} {\bibfnamefont
  {S.}~\bibnamefont {Pancaldi}}, \bibinfo {author} {\bibfnamefont
  {I.}~\bibnamefont {Ferrier-Barbut}}, \bibinfo {author} {\bibfnamefont
  {A.}~\bibnamefont {Browaeys}},\ and\ \bibinfo {author} {\bibfnamefont
  {A.~M.}\ \bibnamefont {Rey}},\ }\href@noop {} {\bibinfo {title} {Directional
  superradiance in a driven ultracold atomic gas in free-space}} (\bibinfo
  {year} {2024}),\ \Eprint {https://arxiv.org/abs/2403.15556} {arXiv:2403.15556
  [cond-mat.quant-gas]} \BibitemShut {NoStop}%
\bibitem [{\citenamefont {Goncalves}\ \emph {et~al.}(2024)\citenamefont
  {Goncalves}, \citenamefont {Bombieri}, \citenamefont {Ferioli}, \citenamefont
  {Pancaldi}, \citenamefont {Ferrier-Barbut}, \citenamefont {Browaeys},
  \citenamefont {Shahmoon},\ and\ \citenamefont {Chang}}]{goncalves2024}%
  \BibitemOpen
  \bibfield  {author} {\bibinfo {author} {\bibfnamefont {D.}~\bibnamefont
  {Goncalves}}, \bibinfo {author} {\bibfnamefont {L.}~\bibnamefont {Bombieri}},
  \bibinfo {author} {\bibfnamefont {G.}~\bibnamefont {Ferioli}}, \bibinfo
  {author} {\bibfnamefont {S.}~\bibnamefont {Pancaldi}}, \bibinfo {author}
  {\bibfnamefont {I.}~\bibnamefont {Ferrier-Barbut}}, \bibinfo {author}
  {\bibfnamefont {A.}~\bibnamefont {Browaeys}}, \bibinfo {author}
  {\bibfnamefont {E.}~\bibnamefont {Shahmoon}},\ and\ \bibinfo {author}
  {\bibfnamefont {D.~E.}\ \bibnamefont {Chang}},\ }\href@noop {} {\bibinfo
  {title} {Driven-dissipative phase separation in free-space atomic ensembles}}
  (\bibinfo {year} {2024}),\ \Eprint {https://arxiv.org/abs/2403.15237}
  {arXiv:2403.15237 [quant-ph]} \BibitemShut {NoStop}%
\bibitem [{\citenamefont {Ruostekoski}(2023)}]{Ruostekoski2023}%
  \BibitemOpen
  \bibfield  {author} {\bibinfo {author} {\bibfnamefont {J.}~\bibnamefont
  {Ruostekoski}},\ }\bibfield  {title} {\bibinfo {title} {Cooperative
  quantum-optical planar arrays of atoms},\ }\href
  {https://doi.org/10.1103/PhysRevA.108.030101} {\bibfield  {journal} {\bibinfo
   {journal} {Phys. Rev. A}\ }\textbf {\bibinfo {volume} {108}},\ \bibinfo
  {pages} {030101} (\bibinfo {year} {2023})}\BibitemShut {NoStop}%
\bibitem [{\citenamefont {Bettles}\ \emph {et~al.}(2020)\citenamefont
  {Bettles}, \citenamefont {Lee}, \citenamefont {Gardiner},\ and\ \citenamefont
  {Ruostekoski}}]{Bettles2020}%
  \BibitemOpen
  \bibfield  {author} {\bibinfo {author} {\bibfnamefont {R.~J.}\ \bibnamefont
  {Bettles}}, \bibinfo {author} {\bibfnamefont {M.~D.}\ \bibnamefont {Lee}},
  \bibinfo {author} {\bibfnamefont {S.~A.}\ \bibnamefont {Gardiner}},\ and\
  \bibinfo {author} {\bibfnamefont {J.}~\bibnamefont {Ruostekoski}},\
  }\bibfield  {title} {\bibinfo {title} {{Quantum and nonlinear effects in
  light transmitted through planar atomic arrays}},\ }\href
  {https://doi.org/10.1038/s42005-020-00404-3} {\bibfield  {journal} {\bibinfo
  {journal} {Commun. Phys.}\ }\textbf {\bibinfo {volume} {3}},\ \bibinfo
  {pages} {141} (\bibinfo {year} {2020})}\BibitemShut {NoStop}%
\bibitem [{\citenamefont {Parmee}\ and\ \citenamefont
  {Ruostekoski}(2020)}]{Parmee2020}%
  \BibitemOpen
  \bibfield  {author} {\bibinfo {author} {\bibfnamefont {C.~D.}\ \bibnamefont
  {Parmee}}\ and\ \bibinfo {author} {\bibfnamefont {J.}~\bibnamefont
  {Ruostekoski}},\ }\bibfield  {title} {\bibinfo {title} {{Signatures of
  optical phase transitions in superradiant and subradiant atomic arrays}},\
  }\href {https://doi.org/10.1038/s42005-020-00476-1} {\bibfield  {journal}
  {\bibinfo  {journal} {Commun. Phys.}\ }\textbf {\bibinfo {volume} {3}},\
  \bibinfo {pages} {205} (\bibinfo {year} {2020})}\BibitemShut {NoStop}%
\bibitem [{\citenamefont {Jackson}(1999)}]{Jackson}%
  \BibitemOpen
  \bibfield  {author} {\bibinfo {author} {\bibfnamefont {J.~D.}\ \bibnamefont
  {Jackson}},\ }\href@noop {} {\emph {\bibinfo {title} {Classical
  Electrodynamics}}},\ \bibinfo {edition} {3rd}\ ed.\ (\bibinfo  {publisher}
  {Wiley, New York},\ \bibinfo {year} {1999})\BibitemShut {NoStop}%
\bibitem [{\citenamefont {Facchinetti}\ \emph {et~al.}(2016)\citenamefont
  {Facchinetti}, \citenamefont {Jenkins},\ and\ \citenamefont
  {Ruostekoski}}]{Facchinetti16}%
  \BibitemOpen
  \bibfield  {author} {\bibinfo {author} {\bibfnamefont {G.}~\bibnamefont
  {Facchinetti}}, \bibinfo {author} {\bibfnamefont {S.~D.}\ \bibnamefont
  {Jenkins}},\ and\ \bibinfo {author} {\bibfnamefont {J.}~\bibnamefont
  {Ruostekoski}},\ }\bibfield  {title} {\bibinfo {title} {Storing light with
  subradiant correlations in arrays of atoms},\ }\href
  {https://doi.org/10.1103/PhysRevLett.117.243601} {\bibfield  {journal}
  {\bibinfo  {journal} {Phys. Rev. Lett.}\ }\textbf {\bibinfo {volume} {117}},\
  \bibinfo {pages} {243601} (\bibinfo {year} {2016})}\BibitemShut {NoStop}%
\bibitem [{\citenamefont {Manzoni}\ \emph {et~al.}(2018)\citenamefont
  {Manzoni}, \citenamefont {Moreno-Cardoner}, \citenamefont {Asenjo-Garcia},
  \citenamefont {Porto}, \citenamefont {Gorshkov},\ and\ \citenamefont
  {Chang}}]{Manzoni18}%
  \BibitemOpen
  \bibfield  {author} {\bibinfo {author} {\bibfnamefont {M.~T.}\ \bibnamefont
  {Manzoni}}, \bibinfo {author} {\bibfnamefont {M.}~\bibnamefont
  {Moreno-Cardoner}}, \bibinfo {author} {\bibfnamefont {A.}~\bibnamefont
  {Asenjo-Garcia}}, \bibinfo {author} {\bibfnamefont {J.~V.}\ \bibnamefont
  {Porto}}, \bibinfo {author} {\bibfnamefont {A.~V.}\ \bibnamefont
  {Gorshkov}},\ and\ \bibinfo {author} {\bibfnamefont {D.~E.}\ \bibnamefont
  {Chang}},\ }\bibfield  {title} {\bibinfo {title} {{Optimization of photon
  storage fidelity in ordered atomic arrays}},\ }\href
  {https://doi.org/10.1088/1367-2630/aadb74} {\bibfield  {journal} {\bibinfo
  {journal} {New J. Phys.}\ }\textbf {\bibinfo {volume} {20}},\ \bibinfo
  {pages} {083048} (\bibinfo {year} {2018})}\BibitemShut {NoStop}%
\bibitem [{\citenamefont {Ruostekoski}\ and\ \citenamefont
  {Javanainen}(1997)}]{Ruostekoski1997a}%
  \BibitemOpen
  \bibfield  {author} {\bibinfo {author} {\bibfnamefont {J.}~\bibnamefont
  {Ruostekoski}}\ and\ \bibinfo {author} {\bibfnamefont {J.}~\bibnamefont
  {Javanainen}},\ }\bibfield  {title} {\bibinfo {title} {Quantum field theory
  of cooperative atom response: Low light intensity},\ }\href
  {https://doi.org/10.1103/PhysRevA.55.513} {\bibfield  {journal} {\bibinfo
  {journal} {Phys. Rev. A}\ }\textbf {\bibinfo {volume} {55}},\ \bibinfo
  {pages} {513} (\bibinfo {year} {1997})}\BibitemShut {NoStop}%
\bibitem [{\citenamefont {Rusek}\ \emph {et~al.}(1996)\citenamefont {Rusek},
  \citenamefont {Or\l{}owski},\ and\ \citenamefont {Mostowski}}]{Rusek96}%
  \BibitemOpen
  \bibfield  {author} {\bibinfo {author} {\bibfnamefont {M.}~\bibnamefont
  {Rusek}}, \bibinfo {author} {\bibfnamefont {A.}~\bibnamefont {Or\l{}owski}},\
  and\ \bibinfo {author} {\bibfnamefont {J.}~\bibnamefont {Mostowski}},\
  }\bibfield  {title} {\bibinfo {title} {Localization of light in
  three-dimensional random dielectric media},\ }\href
  {https://doi.org/10.1103/PhysRevE.53.4122} {\bibfield  {journal} {\bibinfo
  {journal} {Phys. Rev. E}\ }\textbf {\bibinfo {volume} {53}},\ \bibinfo
  {pages} {4122} (\bibinfo {year} {1996})}\BibitemShut {NoStop}%
\bibitem [{\citenamefont {Jenkins}\ \emph
  {et~al.}(2016{\natexlab{b}})\citenamefont {Jenkins}, \citenamefont
  {Ruostekoski}, \citenamefont {Javanainen}, \citenamefont {Jennewein},
  \citenamefont {Bourgain}, \citenamefont {Pellegrino}, \citenamefont
  {Sortais},\ and\ \citenamefont {Browaeys}}]{Jenkins_long16}%
  \BibitemOpen
  \bibfield  {author} {\bibinfo {author} {\bibfnamefont {S.~D.}\ \bibnamefont
  {Jenkins}}, \bibinfo {author} {\bibfnamefont {J.}~\bibnamefont
  {Ruostekoski}}, \bibinfo {author} {\bibfnamefont {J.}~\bibnamefont
  {Javanainen}}, \bibinfo {author} {\bibfnamefont {S.}~\bibnamefont
  {Jennewein}}, \bibinfo {author} {\bibfnamefont {R.}~\bibnamefont {Bourgain}},
  \bibinfo {author} {\bibfnamefont {J.}~\bibnamefont {Pellegrino}}, \bibinfo
  {author} {\bibfnamefont {Y.~R.~P.}\ \bibnamefont {Sortais}},\ and\ \bibinfo
  {author} {\bibfnamefont {A.}~\bibnamefont {Browaeys}},\ }\bibfield  {title}
  {\bibinfo {title} {Collective resonance fluorescence in small and dense atom
  clouds: Comparison between theory and experiment},\ }\href
  {https://doi.org/10.1103/PhysRevA.94.023842} {\bibfield  {journal} {\bibinfo
  {journal} {Phys. Rev. A}\ }\textbf {\bibinfo {volume} {94}},\ \bibinfo
  {pages} {023842} (\bibinfo {year} {2016}{\natexlab{b}})}\BibitemShut
  {NoStop}%
\bibitem [{\citenamefont {Lee}\ \emph {et~al.}(2011)\citenamefont {Lee},
  \citenamefont {H{\"{a}}ffner},\ and\ \citenamefont {Cross}}]{Lee2011}%
  \BibitemOpen
  \bibfield  {author} {\bibinfo {author} {\bibfnamefont {T.~E.}\ \bibnamefont
  {Lee}}, \bibinfo {author} {\bibfnamefont {H.}~\bibnamefont {H{\"{a}}ffner}},\
  and\ \bibinfo {author} {\bibfnamefont {M.~C.}\ \bibnamefont {Cross}},\
  }\bibfield  {title} {\bibinfo {title} {{Antiferromagnetic phase transition in
  a nonequilibrium lattice of Rydberg atoms}},\ }\href
  {https://doi.org/10.1103/PhysRevA.84.031402} {\bibfield  {journal} {\bibinfo
  {journal} {Phys. Rev. A}\ }\textbf {\bibinfo {volume} {84}},\ \bibinfo
  {pages} {031402} (\bibinfo {year} {2011})}\BibitemShut {NoStop}%
\bibitem [{\citenamefont {{\v{S}}ibali{\'{c}}}\ \emph
  {et~al.}(2016)\citenamefont {{\v{S}}ibali{\'{c}}}, \citenamefont {Wade},
  \citenamefont {Adams}, \citenamefont {Weatherill},\ and\ \citenamefont
  {Pohl}}]{Sibalic2016}%
  \BibitemOpen
  \bibfield  {author} {\bibinfo {author} {\bibfnamefont {N.}~\bibnamefont
  {{\v{S}}ibali{\'{c}}}}, \bibinfo {author} {\bibfnamefont {C.~G.}\
  \bibnamefont {Wade}}, \bibinfo {author} {\bibfnamefont {C.~S.}\ \bibnamefont
  {Adams}}, \bibinfo {author} {\bibfnamefont {K.~J.}\ \bibnamefont
  {Weatherill}},\ and\ \bibinfo {author} {\bibfnamefont {T.}~\bibnamefont
  {Pohl}},\ }\bibfield  {title} {\bibinfo {title} {{Driven-dissipative
  many-body systems with mixed power-law interactions: Bistabilities and
  temperature-driven nonequilibrium phase transitions}},\ }\href
  {https://doi.org/10.1103/PhysRevA.94.011401} {\bibfield  {journal} {\bibinfo
  {journal} {Phys. Rev. A}\ }\textbf {\bibinfo {volume} {94}},\ \bibinfo
  {pages} {011401} (\bibinfo {year} {2016})}\BibitemShut {NoStop}%
\bibitem [{\citenamefont {Parmee}\ and\ \citenamefont
  {Cooper}(2018)}]{Parmee2018}%
  \BibitemOpen
  \bibfield  {author} {\bibinfo {author} {\bibfnamefont {C.~D.}\ \bibnamefont
  {Parmee}}\ and\ \bibinfo {author} {\bibfnamefont {N.~R.}\ \bibnamefont
  {Cooper}},\ }\bibfield  {title} {\bibinfo {title} {Phases of driven two-level
  systems with nonlocal dissipation},\ }\href
  {https://doi.org/10.1103/PhysRevA.97.053616} {\bibfield  {journal} {\bibinfo
  {journal} {Phys. Rev. A}\ }\textbf {\bibinfo {volume} {97}},\ \bibinfo
  {pages} {053616} (\bibinfo {year} {2018})}\BibitemShut {NoStop}%
\bibitem [{\citenamefont {Carr}\ \emph {et~al.}(2013)\citenamefont {Carr},
  \citenamefont {Ritter}, \citenamefont {Wade}, \citenamefont {Adams},\ and\
  \citenamefont {Weatherill}}]{Carr2013}%
  \BibitemOpen
  \bibfield  {author} {\bibinfo {author} {\bibfnamefont {C.}~\bibnamefont
  {Carr}}, \bibinfo {author} {\bibfnamefont {R.}~\bibnamefont {Ritter}},
  \bibinfo {author} {\bibfnamefont {C.~G.}\ \bibnamefont {Wade}}, \bibinfo
  {author} {\bibfnamefont {C.~S.}\ \bibnamefont {Adams}},\ and\ \bibinfo
  {author} {\bibfnamefont {K.~J.}\ \bibnamefont {Weatherill}},\ }\bibfield
  {title} {\bibinfo {title} {{Nonequilibrium Phase Transition in a Dilute
  Rydberg Ensemble}},\ }\href {https://doi.org/10.1103/PhysRevLett.111.113901}
  {\bibfield  {journal} {\bibinfo  {journal} {Phys. Rev. Lett.}\ }\textbf
  {\bibinfo {volume} {111}},\ \bibinfo {pages} {113901} (\bibinfo {year}
  {2013})}\BibitemShut {NoStop}%
\bibitem [{\citenamefont {Olmos}\ \emph {et~al.}(2013)\citenamefont {Olmos},
  \citenamefont {Yu}, \citenamefont {Singh}, \citenamefont {Schreck},
  \citenamefont {Bongs},\ and\ \citenamefont {Lesanovsky}}]{Olmos13}%
  \BibitemOpen
  \bibfield  {author} {\bibinfo {author} {\bibfnamefont {B.}~\bibnamefont
  {Olmos}}, \bibinfo {author} {\bibfnamefont {D.}~\bibnamefont {Yu}}, \bibinfo
  {author} {\bibfnamefont {Y.}~\bibnamefont {Singh}}, \bibinfo {author}
  {\bibfnamefont {F.}~\bibnamefont {Schreck}}, \bibinfo {author} {\bibfnamefont
  {K.}~\bibnamefont {Bongs}},\ and\ \bibinfo {author} {\bibfnamefont
  {I.}~\bibnamefont {Lesanovsky}},\ }\bibfield  {title} {\bibinfo {title}
  {Long-range interacting many-body systems with alkaline-earth-metal atoms},\
  }\href {https://doi.org/10.1103/PhysRevLett.110.143602} {\bibfield  {journal}
  {\bibinfo  {journal} {Phys. Rev. Lett.}\ }\textbf {\bibinfo {volume} {110}},\
  \bibinfo {pages} {143602} (\bibinfo {year} {2013})}\BibitemShut {NoStop}%
\bibitem [{\citenamefont {Ballantine}\ \emph {et~al.}(2022)\citenamefont
  {Ballantine}, \citenamefont {Wilkowski},\ and\ \citenamefont
  {Ruostekoski}}]{Ballantine22str}%
  \BibitemOpen
  \bibfield  {author} {\bibinfo {author} {\bibfnamefont {K.~E.}\ \bibnamefont
  {Ballantine}}, \bibinfo {author} {\bibfnamefont {D.}~\bibnamefont
  {Wilkowski}},\ and\ \bibinfo {author} {\bibfnamefont {J.}~\bibnamefont
  {Ruostekoski}},\ }\bibfield  {title} {\bibinfo {title} {Optical magnetism and
  wavefront control by arrays of strontium atoms},\ }\href
  {https://doi.org/10.1103/PhysRevResearch.4.033242} {\bibfield  {journal}
  {\bibinfo  {journal} {Phys. Rev. Res.}\ }\textbf {\bibinfo {volume} {4}},\
  \bibinfo {pages} {033242} (\bibinfo {year} {2022})}\BibitemShut {NoStop}%
\bibitem [{\citenamefont {Beloy}\ \emph {et~al.}(2012)\citenamefont {Beloy},
  \citenamefont {Sherman}, \citenamefont {Lemke}, \citenamefont {Hinkley},
  \citenamefont {Oates},\ and\ \citenamefont {Ludlow}}]{Beloy12}%
  \BibitemOpen
  \bibfield  {author} {\bibinfo {author} {\bibfnamefont {K.}~\bibnamefont
  {Beloy}}, \bibinfo {author} {\bibfnamefont {J.~A.}\ \bibnamefont {Sherman}},
  \bibinfo {author} {\bibfnamefont {N.~D.}\ \bibnamefont {Lemke}}, \bibinfo
  {author} {\bibfnamefont {N.}~\bibnamefont {Hinkley}}, \bibinfo {author}
  {\bibfnamefont {C.~W.}\ \bibnamefont {Oates}},\ and\ \bibinfo {author}
  {\bibfnamefont {A.~D.}\ \bibnamefont {Ludlow}},\ }\bibfield  {title}
  {\bibinfo {title} {Determination of the $5d6s$ ${}^{3}{D}_{1}$ state lifetime
  and blackbody-radiation clock shift in yb},\ }\href
  {https://doi.org/10.1103/PhysRevA.86.051404} {\bibfield  {journal} {\bibinfo
  {journal} {Phys. Rev. A}\ }\textbf {\bibinfo {volume} {86}},\ \bibinfo
  {pages} {051404} (\bibinfo {year} {2012})}\BibitemShut {NoStop}%
\bibitem [{\citenamefont {Jenkins}\ and\ \citenamefont
  {Ruostekoski}(2012)}]{Jenkins2012a}%
  \BibitemOpen
  \bibfield  {author} {\bibinfo {author} {\bibfnamefont {S.~D.}\ \bibnamefont
  {Jenkins}}\ and\ \bibinfo {author} {\bibfnamefont {J.}~\bibnamefont
  {Ruostekoski}},\ }\bibfield  {title} {\bibinfo {title} {Controlled
  manipulation of light by cooperative response of atoms in an optical
  lattice},\ }\href {https://doi.org/10.1103/PhysRevA.86.031602} {\bibfield
  {journal} {\bibinfo  {journal} {Phys. Rev. A}\ }\textbf {\bibinfo {volume}
  {86}},\ \bibinfo {pages} {031602} (\bibinfo {year} {2012})}\BibitemShut
  {NoStop}%
\bibitem [{\citenamefont {Morsch}\ and\ \citenamefont
  {Oberthaler}(2006)}]{Morsch06}%
  \BibitemOpen
  \bibfield  {author} {\bibinfo {author} {\bibfnamefont {O.}~\bibnamefont
  {Morsch}}\ and\ \bibinfo {author} {\bibfnamefont {M.}~\bibnamefont
  {Oberthaler}},\ }\bibfield  {title} {\bibinfo {title} {Dynamics of
  bose-einstein condensates in optical lattices},\ }\href
  {https://doi.org/10.1103/RevModPhys.78.179} {\bibfield  {journal} {\bibinfo
  {journal} {Rev. Mod. Phys.}\ }\textbf {\bibinfo {volume} {78}},\ \bibinfo
  {pages} {179} (\bibinfo {year} {2006})}\BibitemShut {NoStop}%
\bibitem [{\citenamefont {Hannukainen}\ and\ \citenamefont
  {Larson}(2018)}]{Hannukainen2018}%
  \BibitemOpen
  \bibfield  {author} {\bibinfo {author} {\bibfnamefont {J.}~\bibnamefont
  {Hannukainen}}\ and\ \bibinfo {author} {\bibfnamefont {J.}~\bibnamefont
  {Larson}},\ }\bibfield  {title} {\bibinfo {title} {Dissipation-driven quantum
  phase transitions and symmetry breaking},\ }\href
  {https://doi.org/10.1103/PhysRevA.98.042113} {\bibfield  {journal} {\bibinfo
  {journal} {Phys. Rev. A}\ }\textbf {\bibinfo {volume} {98}},\ \bibinfo
  {pages} {042113} (\bibinfo {year} {2018})}\BibitemShut {NoStop}%
\bibitem [{\citenamefont {Carmichael}(1986)}]{Carmichael1986a}%
  \BibitemOpen
  \bibfield  {author} {\bibinfo {author} {\bibfnamefont {H.}~\bibnamefont
  {Carmichael}},\ }\bibfield  {title} {\bibinfo {title} {{"Theory of Quantum
  Fluctuations in Optical Bistability"}},\ }in\ \href@noop {} {\emph {\bibinfo
  {booktitle} {Front. Quantum Opt.}}}\ (\bibinfo  {publisher} {Adam Hilger,
  Bristol},\ \bibinfo {year} {1986})\ pp.\ \bibinfo {pages}
  {120--203}\BibitemShut {NoStop}%
\bibitem [{\citenamefont {Olmos}\ \emph {et~al.}(2014)\citenamefont {Olmos},
  \citenamefont {Yu},\ and\ \citenamefont {Lesanovsky}}]{Olmos2014}%
  \BibitemOpen
  \bibfield  {author} {\bibinfo {author} {\bibfnamefont {B.}~\bibnamefont
  {Olmos}}, \bibinfo {author} {\bibfnamefont {D.}~\bibnamefont {Yu}},\ and\
  \bibinfo {author} {\bibfnamefont {I.}~\bibnamefont {Lesanovsky}},\ }\bibfield
   {title} {\bibinfo {title} {{Steady-state properties of a driven atomic
  ensemble with nonlocal dissipation}},\ }\href
  {https://doi.org/10.1103/PhysRevA.89.023616} {\bibfield  {journal} {\bibinfo
  {journal} {Phys. Rev. A}\ }\textbf {\bibinfo {volume} {89}},\ \bibinfo
  {pages} {023616} (\bibinfo {year} {2014})}\BibitemShut {NoStop}%
\bibitem [{\citenamefont {Rempe}\ \emph {et~al.}(1991)\citenamefont {Rempe},
  \citenamefont {Thompson}, \citenamefont {Brecha}, \citenamefont {Lee},\ and\
  \citenamefont {Kimble}}]{Rempe1991}%
  \BibitemOpen
  \bibfield  {author} {\bibinfo {author} {\bibfnamefont {G.}~\bibnamefont
  {Rempe}}, \bibinfo {author} {\bibfnamefont {R.~J.}\ \bibnamefont {Thompson}},
  \bibinfo {author} {\bibfnamefont {R.~J.}\ \bibnamefont {Brecha}}, \bibinfo
  {author} {\bibfnamefont {W.~D.}\ \bibnamefont {Lee}},\ and\ \bibinfo {author}
  {\bibfnamefont {H.~J.}\ \bibnamefont {Kimble}},\ }\bibfield  {title}
  {\bibinfo {title} {{Optical bistability and photon statistics in cavity
  quantum electrodynamics}},\ }\href
  {https://doi.org/10.1103/PhysRevLett.67.1727} {\bibfield  {journal} {\bibinfo
   {journal} {Phys. Rev. Lett.}\ }\textbf {\bibinfo {volume} {67}},\ \bibinfo
  {pages} {1727} (\bibinfo {year} {1991})}\BibitemShut {NoStop}%
\bibitem [{\citenamefont {Alaeian}\ and\ \citenamefont {Bu{\v
  c}a}(2022)}]{Alaeian2022}%
  \BibitemOpen
  \bibfield  {author} {\bibinfo {author} {\bibfnamefont {H.}~\bibnamefont
  {Alaeian}}\ and\ \bibinfo {author} {\bibfnamefont {B.}~\bibnamefont {Bu{\v
  c}a}},\ }\bibfield  {title} {\bibinfo {title} {Exact multistability and
  dissipative time crystals in interacting fermionic lattices},\ }\href
  {https://doi.org/10.1038/s42005-022-01090-z} {\bibfield  {journal} {\bibinfo
  {journal} {Communications Physics}\ }\textbf {\bibinfo {volume} {5}},\
  \bibinfo {pages} {318} (\bibinfo {year} {2022})}\BibitemShut {NoStop}%
\bibitem [{\citenamefont {Leppenen}\ and\ \citenamefont
  {Shahmoon}(2024)}]{Leppenen2024}%
  \BibitemOpen
  \bibfield  {author} {\bibinfo {author} {\bibfnamefont {N.}~\bibnamefont
  {Leppenen}}\ and\ \bibinfo {author} {\bibfnamefont {E.}~\bibnamefont
  {Shahmoon}},\ }\href@noop {} {\bibinfo {title} {Quantum bistability at the
  interplay between collective and individual decay}} (\bibinfo {year}
  {2024}),\ \Eprint {https://arxiv.org/abs/2404.02134} {arXiv:2404.02134
  [quant-ph]} \BibitemShut {NoStop}%
\bibitem [{\citenamefont {Javanainen}\ and\ \citenamefont
  {Rajapakse}(2019)}]{Javanainen19}%
  \BibitemOpen
  \bibfield  {author} {\bibinfo {author} {\bibfnamefont {J.}~\bibnamefont
  {Javanainen}}\ and\ \bibinfo {author} {\bibfnamefont {R.}~\bibnamefont
  {Rajapakse}},\ }\bibfield  {title} {\bibinfo {title} {Light propagation in
  systems involving two-dimensional atomic lattices},\ }\href
  {https://doi.org/10.1103/PhysRevA.100.013616} {\bibfield  {journal} {\bibinfo
   {journal} {Phys. Rev. A}\ }\textbf {\bibinfo {volume} {100}},\ \bibinfo
  {pages} {013616} (\bibinfo {year} {2019})}\BibitemShut {NoStop}%
\bibitem [{\citenamefont {Carmichael}\ and\ \citenamefont
  {Kim}(2000)}]{carmichael2000}%
  \BibitemOpen
  \bibfield  {author} {\bibinfo {author} {\bibfnamefont {H.}~\bibnamefont
  {Carmichael}}\ and\ \bibinfo {author} {\bibfnamefont {K.}~\bibnamefont
  {Kim}},\ }\bibfield  {title} {\bibinfo {title} {A quantum trajectory
  unraveling of the superradiance master equation},\ }\href
  {https://doi.org/10.1016/S0030-4018(99)00694-X} {\bibfield  {journal}
  {\bibinfo  {journal} {Opt. Commun.}\ }\textbf {\bibinfo {volume} {179}},\
  \bibinfo {pages} {417} (\bibinfo {year} {2000})}\BibitemShut {NoStop}%
\end{thebibliography}
\end{document}